\documentclass[aps,prl,twocolumn,superscriptaddress,longbibliography]{revtex4-2}

\usepackage{graphicx}  
\usepackage{dcolumn}   
\usepackage{bm}        
\usepackage{amssymb}   
\usepackage{amsmath, latexsym}
\usepackage{units}
\usepackage[utf8]{inputenc}
\usepackage[dvipsnames]{xcolor}
\usepackage[%
colorlinks=true,
urlcolor=blue,
linkcolor=blue,
citecolor=blue
]{hyperref}
\usepackage{etoolbox}

\usepackage[type1]{libertine}           
\usepackage{textcomp}
\usepackage[scaled=.85]{beramono}
\usepackage[libertine,cmintegrals,cmbraces,vvarbb,slantedGreek]{newtxmath}
\usepackage[scr=boondoxo]{mathalfa}
\usepackage{bm}
\usepackage[lf]{carlito}

\makeatletter
\DeclareRobustCommand{\cev}[1]{%
  \mathpalette\do@cev{#1}%
}
\newcommand{\do@cev}[2]{%
  \fix@cev{#1}{+}%
  \reflectbox{$\m@th#1\vec{\reflectbox{$\fix@cev{#1}{-}\m@th#1#2\fix@cev{#1}{+}$}}$}%
  \fix@cev{#1}{-}%
}
\newcommand{\fix@cev}[2]{%
  \ifx#1\displaystyle
    \mkern#23mu
  \else
    \ifx#1\textstyle
      \mkern#23mu
    \else
      \ifx#1\scriptstyle
        \mkern#22mu
      \else
        \mkern#22mu
      \fi
    \fi
  \fi
}
\makeatother

\newcommand{\figref}[1]{Fig.\,\ref{#1}}
\renewcommand{\eqref}[1]{Eq.\,(\ref{#1})}
\newcommand{\iv}{$I(V)$}
\newcommand{\iz}{$I(z)$}

\begin{document}

\title{Extracting the Transport Channel Transmissions in Scanning Tunneling Microscopy using the Superconducting Excess Current}

\author{Jacob Senkpiel}
\affiliation{Max-Planck-Institut f\"ur Festk\"orperforschung, Heisenbergstraße 1,
	70569 Stuttgart, Germany}
\author{Robert Drost}
\affiliation{Max-Planck-Institut f\"ur Festk\"orperforschung, Heisenbergstraße 1,
	70569 Stuttgart, Germany}
\author{Jan C. Klöckner}
\affiliation{Okinawa Institute of Science and Technology Graduate University, Onna-son, Okinawa 904-0495, Japan}
\affiliation{Fachbereich Physik, Universität Konstanz, 78457 Konstanz, Germany}
\author{Markus Etzkorn}
\affiliation{Institut für Angewandte Physik, TU Braunschweig, Mendelssohnstraße 2, 38106 Braunschweig, Germany}
\author{Joachim Ankerhold}
\affiliation{Institute for Complex Quantum Systems University of Ulm, Albert-Einstein-Allee 11, 89069 Ulm, Germany}
\author{Juan Carlos Cuevas}
\affiliation{Departamento de Física Teórica de la Materia Condensada and Condensed Matter Physics Center (IFIMAC), Universidad Autónoma de Madrid, 28049 Madrid, Spain}
\author{Fabian Pauly}
\affiliation{Institute of Physics, University of Augsburg, 86135 Augsburg, Germany}
\author{Klaus Kern}
\affiliation{Max-Planck-Institut f\"ur Festk\"orperforschung, Heisenbergstraße 1,
	70569 Stuttgart, Germany}
\affiliation{Institut de Physique, École Polytechnique Fédérale de Lausanne, 1015 Lausanne, Switzerland}
\author{Christian R. Ast}
\email[Corresponding author; electronic address:\ ]{c.ast@fkf.mpg.de}
\affiliation{Max-Planck-Institut f\"ur Festk\"orperforschung, Heisenbergstraße 1,
	70569 Stuttgart, Germany}

\date{\today}

\begin{abstract}
Transport through quantum coherent conductors, like atomic junctions, is described by the distribution of conduction channels. Information about the number of channels and their transmission can be extracted from various sources, such as multiple Andreev reflections, dynamical Coulomb blockade, or shot noise. We complement this set of methods by introducing the superconducting excess current as a new tool to continuously extract the transport channel transmissions of an atomic scale junction in a scanning tunneling microscope. In conjunction with \textit{ab initio} simulations, we employ this technique in atomic aluminum junctions to determine the influence of the structure adjacent to the contact atoms on the transport properties.
\end{abstract}
\maketitle

At the heart of the modern theory of quantum transport lies the concept of transport channels \cite{landauer1957spatial, Landauer1970}. Similar to transverse electromagnetic modes in an optical waveguide, charge can be transported between two baths through a set of distinct pathways arising from individual quantum states. These quantum states are rooted in the electronic structure of a device and can be manipulated by changes in materials or geometry \cite{Cuevas1999}. Each transport channel is further characterised by a transmission value, which specifies the probability for an electron that enters the channel from one bath to be transmitted through it to reach the other bath. The number of transport channels and their respective transmissions, often referred to as the mesoscopic PIN code (in analogy to the personal identification number), characterises a transport configuration.

Established techniques to measure the mesoscopic PIN code are either technically challenging (shot noise measurements) \cite{Brom1999, schoelkopf1997, Blanter2000, Cron2001a, Agrait2003, Djukic2006, Kumar2013, vardimon2013, burtzlaff2015shot, Vardimon2015, Vardimon2016}, time consuming (dynamical Coulomb blockade \cite{senkpiel2020dynamical} or multiple Andreev reflection measurements \cite{Muller1992, Scheer1997, Scheer1998, Chauvin2007, Rocca2007, ludoph2000multiple, riquelme2005distribution,Massee2018,Massee2019,bastiaans_amplifier_2018}), or only applicable to ensemble averages (conductance fluctuations \cite{ludoph1999evidence, ludoph2000conductance}). It is thus difficult to track the continuous evolution of the PIN code as a function of an experimental control parameter to gain deeper insights into the transport process.

In this article, we show that, if at least one of the electrodes involved in transport is superconducting, the excess current can be used to determine the PIN code in a simple and rapid fashion. Using an ultra-low temperature scanning tunnelling microscope (STM), we study transport properties of superconducting tunnel junctions from the deep tunnelling regime to atomic contact and extract the PIN code as a continuous function of the junction conductance. By using \textit{ab initio} transport calculations, we are able to elucidate the microscopic nature of the conduction channels.

In addition to quasi-particle tunnelling, superconducting contacts support another mode for charge transport through Andreev reflections \cite{andreev1964thermal}. An electron incident onto the superconductor is reflected as a hole, thereby transferring a charge of $2e$ into the superconductor and forming a Cooper pair. Higher orders of this process occur in superconductor-superconductor junctions and are referred to as multiple Andreev reflections (MARs).

Even at bias voltages $V$ outside of the superconducting gap ($eV \gg 2\Delta$), the lowest order Andreev reflection continues to contribute to the current leading to a constant offset from the expected single particle current \cite{Blonder1982}. Formally, this excess current is defined as
\begin{equation}
    I_{\text{Exc}} = I_{\text{S}}(V)-I_{\text{N}}(V)\big|_{eV\gg 2\Delta},
    \label{eq:Iexc}
\end{equation}
\noindent where $\Delta$ is the superconducting gap parameter, the subscripts S and N refer to the superconducting and normal state, respectively, and $e$ is the elementary charge. A non-linear dependence of the excess current on the channel transmission allows an extraction of the junction PIN code. Assuming $\Delta_{\text{Tip}} = \Delta_{\text{Sample}} = \Delta$, the excess current across a superconductor-insulator-superconductor junction at zero temperature can be calculated as
\begin{multline}
	 I_\text{Exc} = \dfrac{2 e \upDelta}{h} \sum_i \frac{\tau^2_i}{1 - \tau_i} \times \\ \times \left[ 1 - \frac{\tau^2_i}{2 (2 - \tau_i) \sqrt{1 - \tau_i}} \ln \left( \frac{1 + \sqrt{1 - \tau_i}}{1 - \sqrt{1 - \tau_i}} \right) \right],
	\label{eq:excessCurrentExact}
\end{multline}

\begin{figure}
	\centering
	\includegraphics[width = \columnwidth]{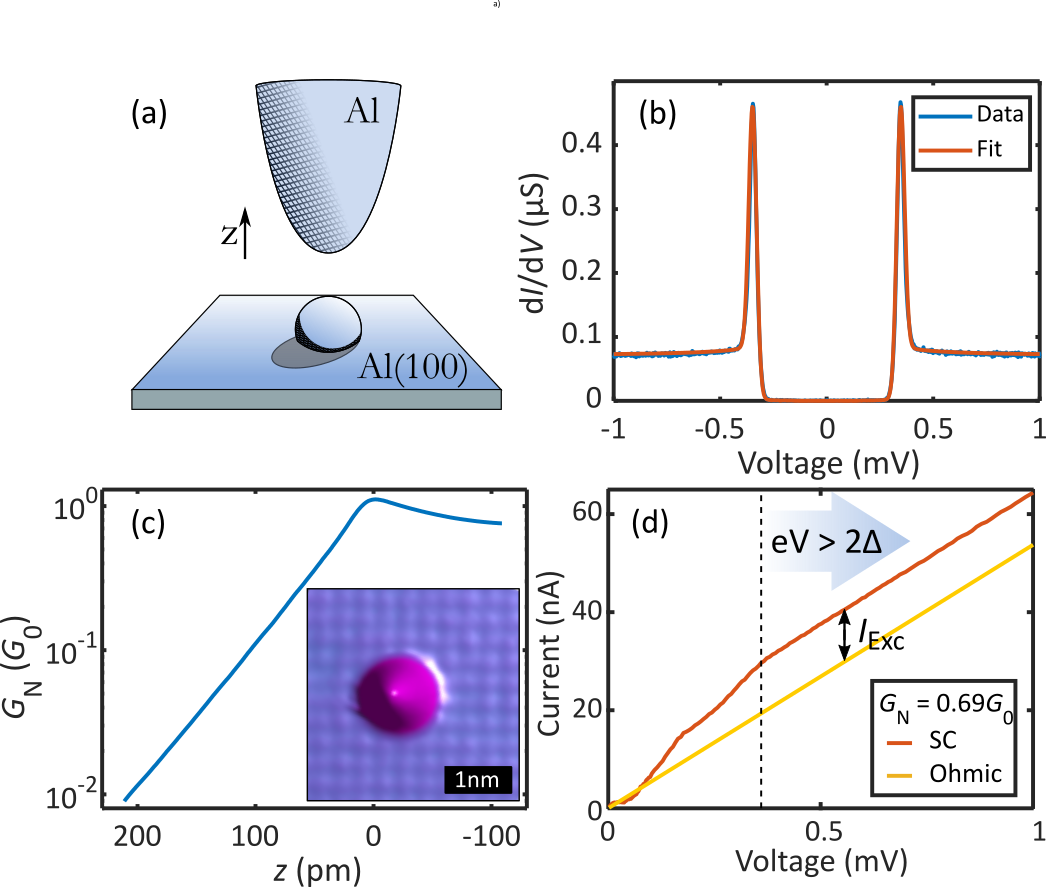}
	\caption{(a) Sketch of the single-atom junction studied in the experiment. The tip height $z$ is adjusted in the course of the experiment. (b) Quasiparticle spectrum at low setpoint conductance (blue) with corresponding fit. (c) Conductance curve $G_\text{N}$($z$) recorded above an Al adatom, showing a clear initial exponential increase of the conductance (setpoint 1.5\,mV at 1\,nA). Inset: Topographic image of an Al adatom on the Al(100) surface. (d) \iv-curve of the superconducting sample at $G_\text{N} = 0.69\,G_0$. The excess current is the $y$-axis intercept of the curve at $eV > 2\Delta$. The dashed vertical line indicates $2\Delta = 360\,\upmu$eV.}
	\label{fig:junction}
\end{figure}

\noindent where $\tau_i$ is the transmission of the $i^{\text{th}}$ electronic transport channel and $h$ is Planck's constant \cite{Cuevas1999}. We assume that the $\tau_i$ are independent of each other and of the bias voltage $V$, a good approximation in the small energy window around zero bias relevant to transport in atomic contacts. As can be seen from \eqref{eq:excessCurrentExact}, the excess current depends on the $\tau_i$ and $\Delta$, but not on the bias voltage $V$ applied to the contact. If only few open transport channels are present, a single data point is thus enough to fully determine the PIN code, greatly facilitating and expediting data acquisition.

All measurements are performed in a custom-built STM placed in a dilution refrigerator and operating at a base temperature of 10\,mK \cite{Assig2013}. The Al(100) sample was cleaned by bombardment with Ar ions followed by stepwise annealing in UHV from 480$^{\circ}$C to 460$^{\circ}$C to 435$^{\circ}$C. We extracted individual Al atoms from the substrate and placed them on the pristine Al(100) surface to create a simple transport configuration which serves as a model system for our PIN code analysis\,\cite{si}.

We approach the adatom with the STM tip to measure conductance spectra at constant height and \iz-curves. The situation is schematically shown in \figref{fig:junction}(a). Conductance spectra acquired at low conductance show the characteristic signature of superconductor-superconductor tunnelling with an energy gap around zero bias, flanked by coherence peaks on either side (see \figref{fig:junction}(b)). The values of $\Delta$ for tip and sample, required for a quantitative analysis based on \eqref{eq:excessCurrentExact} or numerical simulations, can be extracted from a fit (see Supplemental Material for details). For the present case, we find $\Delta = 180\,\upmu$eV.

Junctions such as the one described here have been reported to exhibit exceptional stability \cite{senkpiel2020single}. \figref{fig:junction}(c) shows the height-dependent normal-state conductance $G_\text{N}$ of a typical junction calculated from an \iz-measurement at 1.5\,mV, outside the gap, in units of the quantum of conductance, $G_0 = 2e^2/h$. We define the point $z = 0$ to be at the maximum of conductance. At first, the conductance increases exponentially at a rate of roughly one order of magnitude per 100\,pm, as expected from theory, until reaching a maximum value near $G_0$. If the tip is approached further, the conductance decreases again at a rate which depends on the microtip. In some cases, a reduction to about 0.2\,$G_0$ has been observed. Remarkably, no jump to contact occurs \cite{untiedt2007formation} and there is no hysteresis in the current when retracting the tip \cite{si}. We conclude that the junction remains unchanged after the approach-retract cycle.

While the general shape of the height-dependent conductance is consistent across measurements with different microtips, the value of the maximum conductance $G^{\text{max}}_\text{N}$ attained and the magnitude of the drop past $G^{\text{max}}_\text{N}$ varies significantly\,\cite{si}. These differences must be rooted in the details of quantum transport between tip and sample, as characterized by the mesoscopic PIN code. Monitoring the PIN code continuously as a function of $z$ promises deeper insights into the quantum properties of the junctions.

\begin{figure}
	\centering
	\includegraphics[width = \columnwidth]{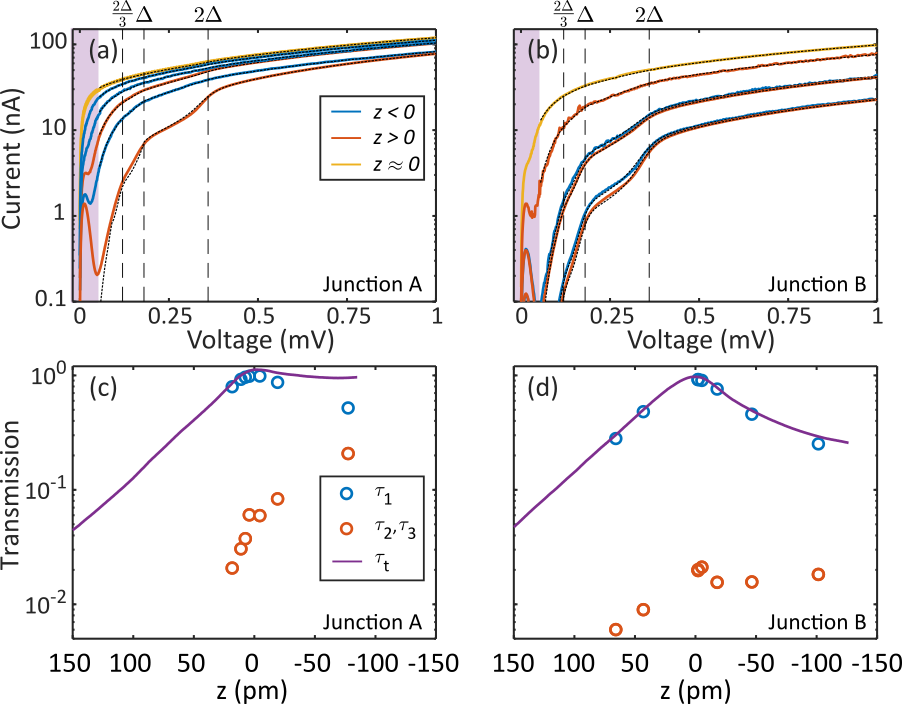} 
	\caption{(a), (b) \iv-curves at various conductances for two typical Al adatoms. MARs appear as steps in the current at fractions of $2\Delta$. Fits to the data are superimposed in black dotted lines. (c), (d) PIN code analysis from the MAR model for the junctions in panels (a) and (b), respectively, as circles. The purple curve shows the total transmission, i.e.\ the conductance $G_\text{N}(z)$ in units of $G_0$.}
	\label{fig:AndreevCurves}
\end{figure}

As can be seen from \eqref{eq:excessCurrentExact}, the complete PIN code is contained in the excess current. The \iv-characteristic of a superconducting tunnel junction evolves to first order in $V$ for $eV > 2\Delta$, but it does not pass through the origin. It instead intersects the current axis at a constant value, referred to as the excess current (see \figref{fig:junction}(d)). It arises from the lowest-order Andreev reflection, which contributes to the total current even outside the gap. Indeed, an established method of extracting the PIN code in superconducting junctions is to analyse the subharmonic gap structure due to MARs in a junction at high conductance. MARs lead to a series of features at integer fractions of $2\Delta$ which characterize the channel configuration. The excess current, having the same physical origin, contains identical information.

To show that an analysis of the excess current is a pertinent way of determining the junction PIN code, we supplement our continuous \iz-measurements of the excess current with full MAR spectra at selected points in the \iz-curve. Representative data sets from two distinct junctions are shown in \figref{fig:AndreevCurves}(a) and (b). The superconducting gap is visible as a step in the \iv-curve at $2\Delta=360\,\upmu$eV, while the MARs are manifested as shoulders at $eV = 2\Delta/2,\; 2\Delta/3,\; ...$. In addition to the MARs, the Josephson effect, the coherent tunnelling of Cooper pairs, is visible as a sharp rise in current close to zero bias. Strong $z$-dependent variations of the sub-gap structure are clearly apparent in both data sets, pointing towards changes in the channel configuration.

\begin{figure}
	\centering
	\includegraphics[width = \columnwidth]{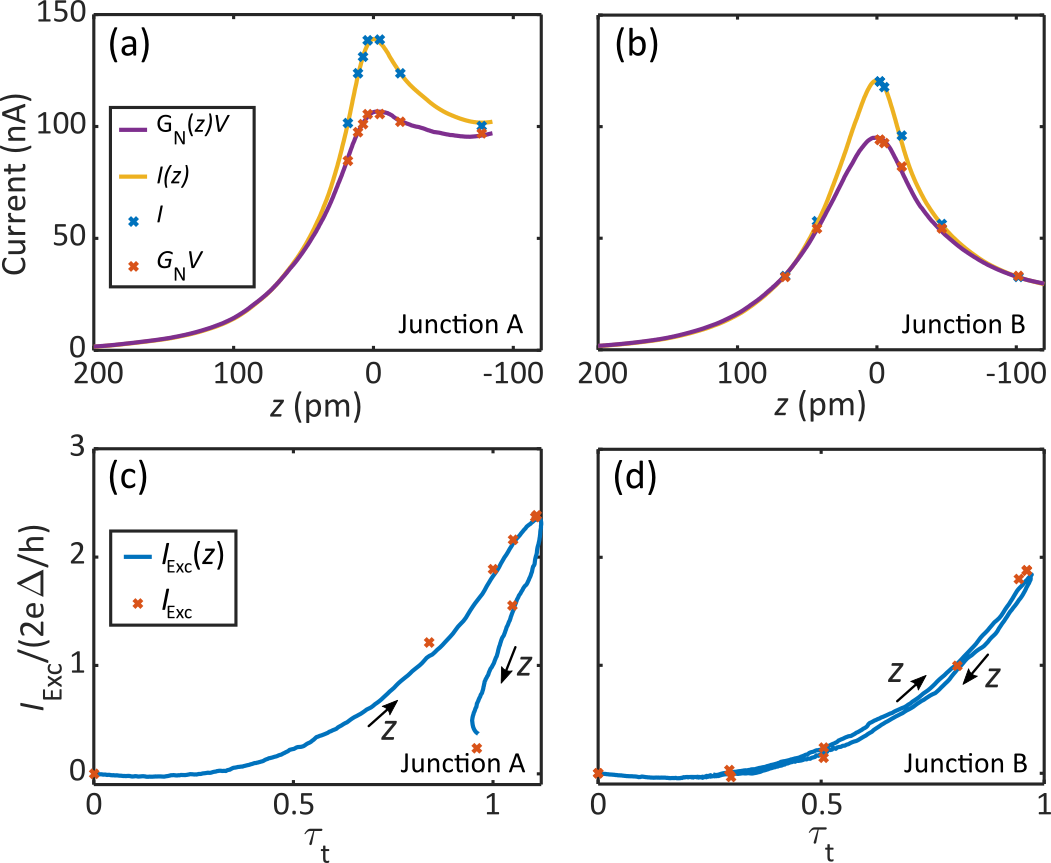}
	\caption{(a), (b) Measured current (yellow) and reconstructed $G_\text{N}(z)V$ (purple) signals for junctions A and B, respectively. The difference between the two curves results from the excess current. The blue and orange crosses show the same quantities from the MAR point spectra in \figref{fig:AndreevCurves}. (c), (d) Excess current as a function of total transmission $\tau_\text{t}$ for junctions A and B, respectively. Note that the curves show the approach curve of the tip, i.e.\ the arrows indicate the direction of decreasing $z$.}
	\label{fig:Iexc}
\end{figure}

We use a well established model based on the relation between MARs and PIN code \cite{Cuevas1996, Cuevas1998} to analyze the channel configuration of the different junctions and their transmission dependence. Owing to its partially filled $p$-shell, we assume that an atomic contact of Al may sustain up to three distinct transport channels. Due to this construction, we find the two transport channels having $\pi$-symmetry to be degenerate. We fit the MAR data using values for $\Delta$ and the Dynes broadening parameter \cite{Dynes1978a}, which we both extract from a quasiparticle fit such as the one shown in \figref{fig:junction}(b) and capture the environmental broadening by a convolution with a Gaussian \cite{senkpiel2020single}. The voltage range between $\pm50\upmu\text{V}$, dominated by Josephson transport, is excluded from the analysis. The results of this channel analysis are presented in \figref{fig:AndreevCurves}(c) and (d). The non-linear dependence of the different orders of Andreev reflections on the channel transmission allows us to extract the junction PIN code. Comparing this with the total conductance, which is well reproduced by our calculations, we confirm the predominant single-channel nature of the contacts.

We now turn toward the task of extracting the excess current from the \iz-traces. As can be seen from \figref{fig:junction}(d), the excess current may be regarded as an integration constant when calculating the current from conductance. We define the normal state current at bias voltage $V$ as
\begin{equation}
    I_{\text{N}}(z) = G_{\text{N}}(z) V,
    \label{eq:normalStateCurrent}
\end{equation}
\noindent where $G_{\text{N}}$ is the differential conductance which is measured directly by means of a lock-in amplifier (see Supplemental Material for details). The excess current is then the difference between the experimentally detected current and the normal state current, see \eqref{eq:Iexc}.

Representative results from the excess current determination are shown in \figref{fig:Iexc} for the same two junctions which have been discussed in the context of \figref{fig:AndreevCurves}. Figure \ref{fig:Iexc}(a) and (b) show the experimentally measured current in yellow and the calculated normal state current according to \eqref{eq:normalStateCurrent} in purple. The blue and orange markers show the same quantities as derived from the spectra shown in \figref{fig:AndreevCurves}(a) and (b). The excess current itself is plotted in \figref{fig:Iexc}(c) and (d). The behaviour of $I_{\text{Exc}}$ is starkly different for both junctions. While $I_{\text{Exc}}$ decreases sharply past the point of maximum conductance in junction A, it is nearly symmetric around the point of maximum conductance in juntion B. We shall now show that the source of these differences lies in the height-dependent evolution of the PIN codes of both junctions.

For speed and convenience, we build a look-up table based on \eqref{eq:excessCurrentExact} containing the excess current at a large number of transmission values and find the best point-by-point match to the \iz-curve using the same constrains as in the MAR analysis above.

\begin{figure}
	\centering
	\includegraphics[width = \columnwidth]{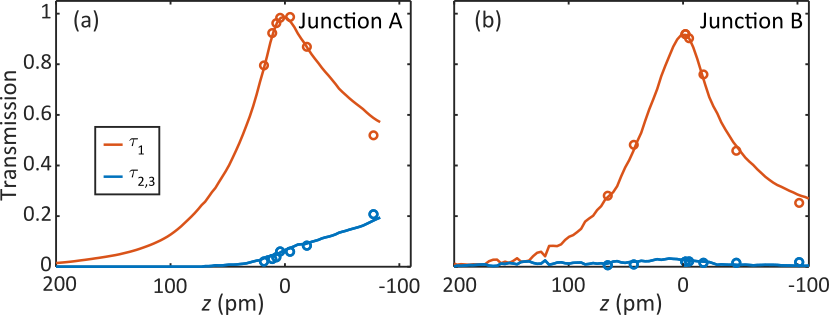}
	\caption{(a), (b) Mesoscopic PIN code analysis using the excess current for junctions A and B, respectively, with the dominant channel in orange and the degenerate channels shown in blue. The circles show the result of the full MAR analysis from \figref{fig:AndreevCurves}.}
	\label{fig:IexcPIN}
\end{figure}

\figref{fig:IexcPIN} shows the resulting transmissions $\tau_i$ for the dominant (orange) and degenerate (blue) channels for junctions A and B. As a control, the results of the full MAR analysis are superimposed as open circles. The analysis of the excess current finds virtually the same channel configuration as the established MAR analysis at those points where data from both methods is available. This agreement is expected since the physical processes at the heart of both methods are the same. The measurement can, hence, be simplified from a full \iv-characteristic to a single data point with no apparent loss of accuracy.

Owing to the ease and speed of the measurement, the excess current provides a very refined picture of the channel evolution as a function of $z$. In general, the dominant channel increases as the tip is approached until reaching nearly unity transmission, at which point $\tau_1$ drops sharply. The transparency of the secondary channels, when present, increases rather monotonously for smaller $z$. A similar drop of the total transmission has been observed in Al junctions before and was attributed to varying channel transmissions under elastic deformation of the contact \cite{cuevas1998a}. A comparatively detailed decomposition into individual transport channels as we report here has not been achieved until now.

\begin{figure}
	\centering
	\includegraphics[width = \columnwidth]{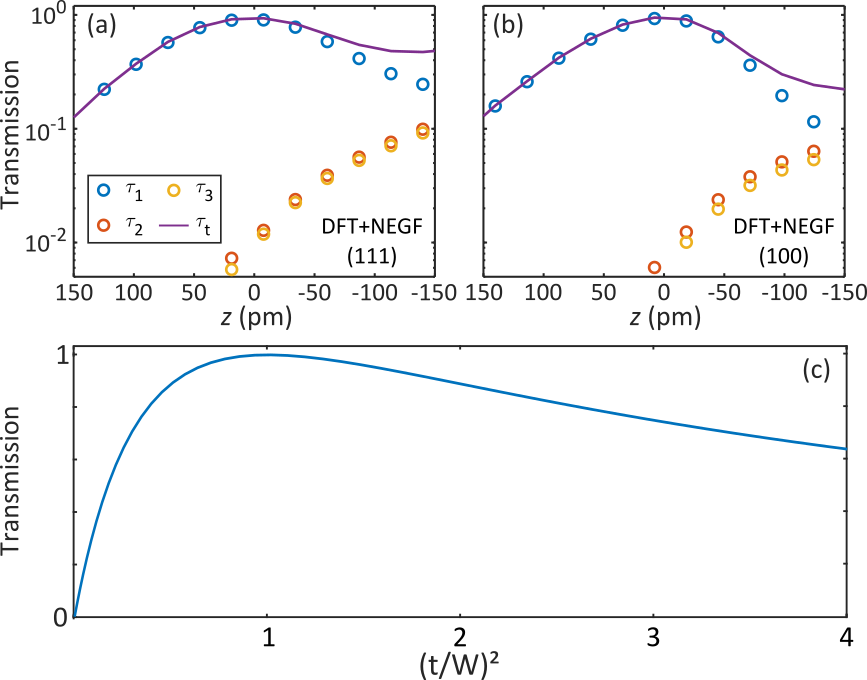}
	\caption{(a), (b) Calculated PIN code for a (111) and (100) terminated tip, respectively, above an Al adatom on Al(100) as a function of tip-sample separation. (c) Transmission through two coupled chains according to \eqref{eq:toyModel}. The model qualitatively reproduces the experimental data.}
	\label{fig:DFT}
\end{figure}

To understand the observed PIN code variations between different junctions and tie them to microscopic origins, we performed \textit{ab initio} simulations using density functional theory (DFT) for two microtips terminated by a (100) and (111) facet, facing a Al(100) surface with a tip atom on top. The junction geometries are optimized at each tip height, and their transport characteristics, including the transmission channels, are computed from the electronic structure through non-equilibrium Green’s function (NEGF) techniques \cite{Pauly2008, si}. The PIN codes obtained from the simulations of the two different tips are shown in \figref{fig:DFT}(a) and (b).

The simulations qualitatively reproduce the experimental observations, with the total conductance reaching a maximum value and subsequently decreasing upon closer approach. The magnitude of the drop-off is dependent on the crystal facet exposed by the tip. As it is unlikely that the tip apex in the experiment is perfectly crystalline, an exact reproduction of the experimental data is not expected. It is clear, though, that changes in PIN code evolution can be traced to structural properties of the microtip.

The DFT simulations also allow us to extract the coupling strength between the tip and sample electrodes, which rises continuously as a function of $z$. The rise does not necessarily translate into a higher transmission, though. This behaviour can be qualitatively understood in a one-dimensional tight-binding model. We thus study two semi-infinite atomic chains with uniform nearest-neighbour hopping amplitude $t_0$ and coupled to each other by the hopping $t$. In the limit $t \ll t_0$, itinerant charges in the chain are likely to be back-scattered from the point of contact. In the other limiting case, $t \gg t_0$, electrons are likely to pass back and forth between the two sides of the junction, thereby blocking it for transport. This kind of system has been described theoretically in Ref.\,\cite{Cuevas1999}. The transmission between the two chains is
\begin{equation}
    \tau = \dfrac{4t^2/W^2}{(1+t^2/W^2)^2},
    \label{eq:toyModel}
\end{equation}
where $t$ is the coupling between the two leads and $W=1/(\pi\rho(E_F))$ is an energy scale related to the density of states at the Fermi level. This model allows us to understand the essential physics of single atom tunnel junctions in a minimal setting. In the weak coupling limit, the small inter-chain hopping acts as a potential barrier that limits the charge transfer between the two leads. However, when the inter-chain coupling becomes larger than the intra-chain hopping, a bound state forms between the leads, which also inhibits charge transfer. These results are in qualitative agreement with our experimental results, see \figref{fig:DFT}(c).

The PIN code is the central quantity for understanding quantum transport in mesoscopic contacts. It is a general property of a transport configuration and does not change between the superconducting and normal states. When at least one of the electrodes participating in transport is superconducting, the PIN code can be derived from Andreev processes. The excess current has its origin in MARs, which continue to contribute to the total current even as $eV >2\Delta$. It, therefore, contains the same physical information as the sub-gap structure, while being both easier and faster to measure.

In summary, we use the excess current to measure the continuous evolution of the PIN code in tunnel junctions. Control experiments using a full MAR analysis confirm the accuracy of our measurements. The ability to extract complex quantum properties such as the PIN code from such a simple measurement hold the promise of understanding transport as a function of an external control parameter, or to relate atomic structure to transport in much greater detail than hitherto possible.

This work was funded in part by the ERC Consolidator Grant AbsoluteSpin (Grant No. 681164). J.C.K. and F.P. thank the Collaborative Research Center (SFB) 767 of the German Research Foundation (DFG) as well as the Okinawa Institute of Science and Technology (OIST) Graduate University for financial support. Part of the numerical modeling was performed using the computational resources of the bwHPC program, namely the bwUniCluster and the JUSTUS HPC facility. J.C.C. acknowledges funding from the Spanish MINECO (Grant No. FIS2017-84057-P).

\clearpage
\newpage

\onecolumngrid
\begin{center}
\textbf{\large Supplementary Material for \\ Extracting the Transport Channel Transmissions in Scanning Tunneling Microscopy using the Superconducting Excess Current}
\end{center}
\vspace{1cm}
\twocolumngrid

\setcounter{figure}{0}
\setcounter{table}{0}
\setcounter{equation}{0}
\renewcommand{\thefigure}{S\arabic{figure}}
\renewcommand{\thetable}{S\Roman{table}}
\renewcommand{\theequation}{S\arabic{equation}}

\section{Sample preparation}
The Al(100) sample was prepared by cycles of Ar-ion sputtering and annealing. The tip was cut from 1\,mm diameter aluminum wire (purity of $99.9999\,\%$) and sputtered in vacuum to remove the surface oxide. The tip was further prepared by field emission in proximity of the Al(100) surface and repeated indentations into the substrate until achieving an atomically sharp topography. All measurements were performed at the base temperature of the STM system of 15\,mK.

\begin{figure}
	\centering
	\includegraphics[width = 0.95\columnwidth]{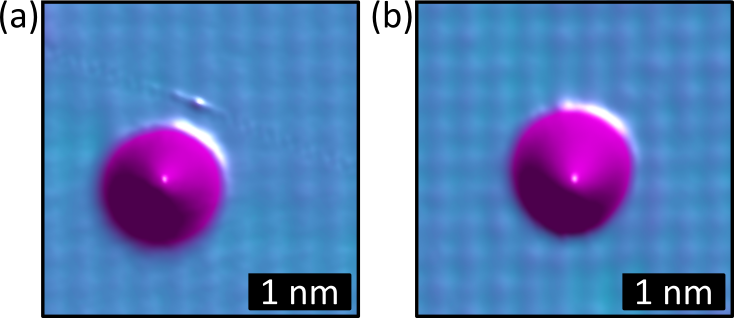}
	\caption{Topographies of the Al adatoms of junctions A (a) and B (b) in the main text.}
	\label{fig:SIAdatoms}
\end{figure}

We developed an experimental procedure, allowing us to reproducibly extract single Al atoms from the surface. The tip is first positioned at the height corresponding to 10\,mV bias and 100\,pA tunnelling current. After turning off the feedback loop, we approach the tip by $-400$\,pm towards the pristine surface and retract it back to the reference height. Following this approach-retract procedure, we observe a single vacancy defect at the point of approach and an individual Al adatom nearby. The adatom can then be moved to a desired position on the surface by atom manipulation. \figref{fig:SIAdatoms} shows the two Al adatoms of junctions A and B from the main text, which we produced by this technique.

\section{Absence of Hysteresis}

\begin{figure}
    \centering
    \includegraphics[width=0.85\columnwidth]{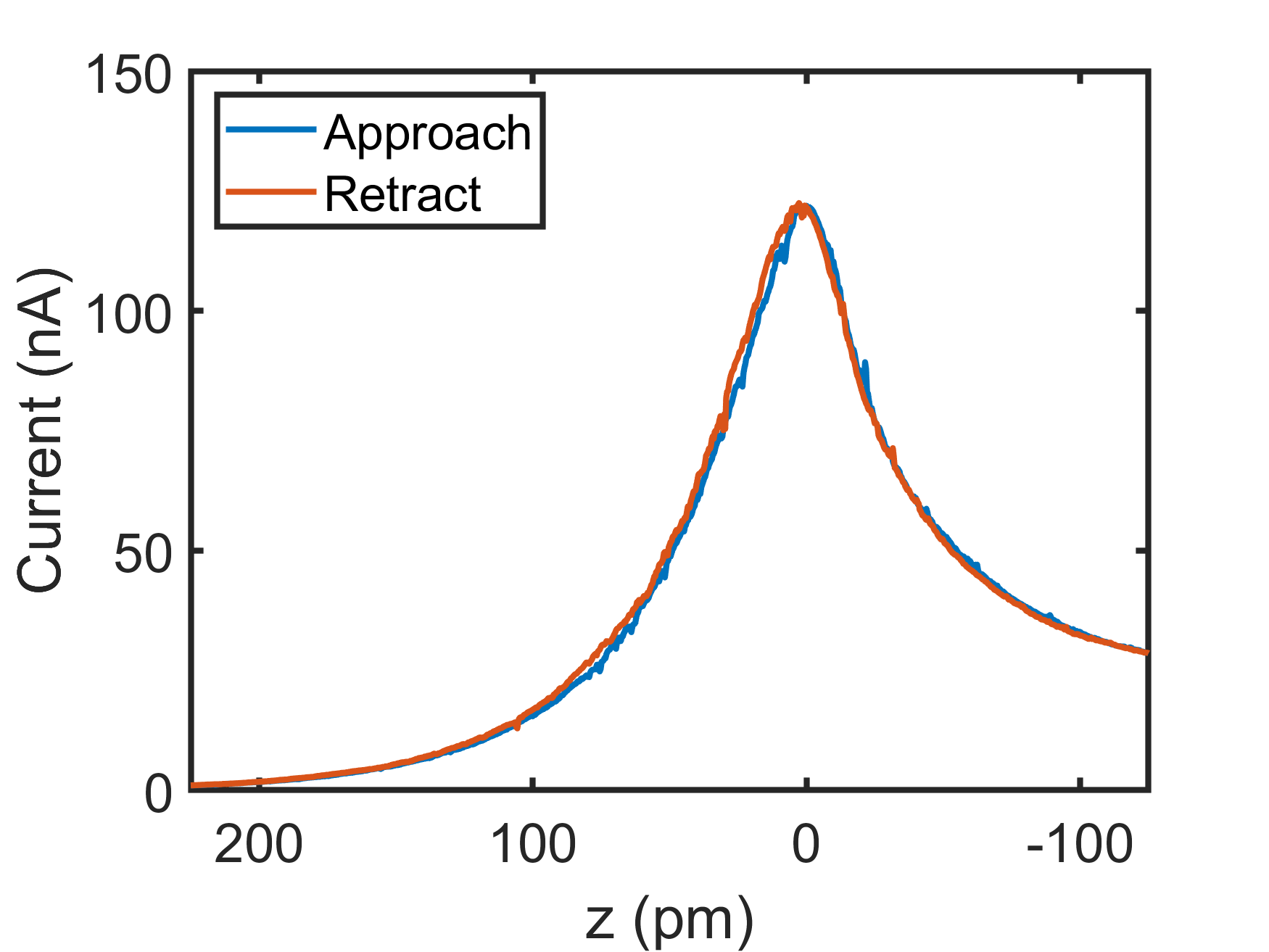}
    \caption{Typical approach-retract curve above an Al adatom. The tip was placed at the reference position of 10\,pA current at 15\,mV bias. After disengaging the feedback loop, the tip is pushed -350\,pm toward the surface and retracted to the starting position. There is no jump to contact and no hysteresis between the approach and retract paths.}
    \label{fig:hysteresisCurrent}
\end{figure}

The single atom junctions we investigated are exceedingly stable and remain unchanged during the approach-retract experiments. There is no jump to contact in the $z$-range explored in our experiment, up to ca. $-100$\,pm past the point of maximum conductance. We observe no hysteresis between the approach and retraction of the tip. An example of a typical approach-retract curve is shown in \figref{fig:hysteresisCurrent}.

\section{Additional Data}

\begin{figure}
	\centering
	\includegraphics[width = \columnwidth]{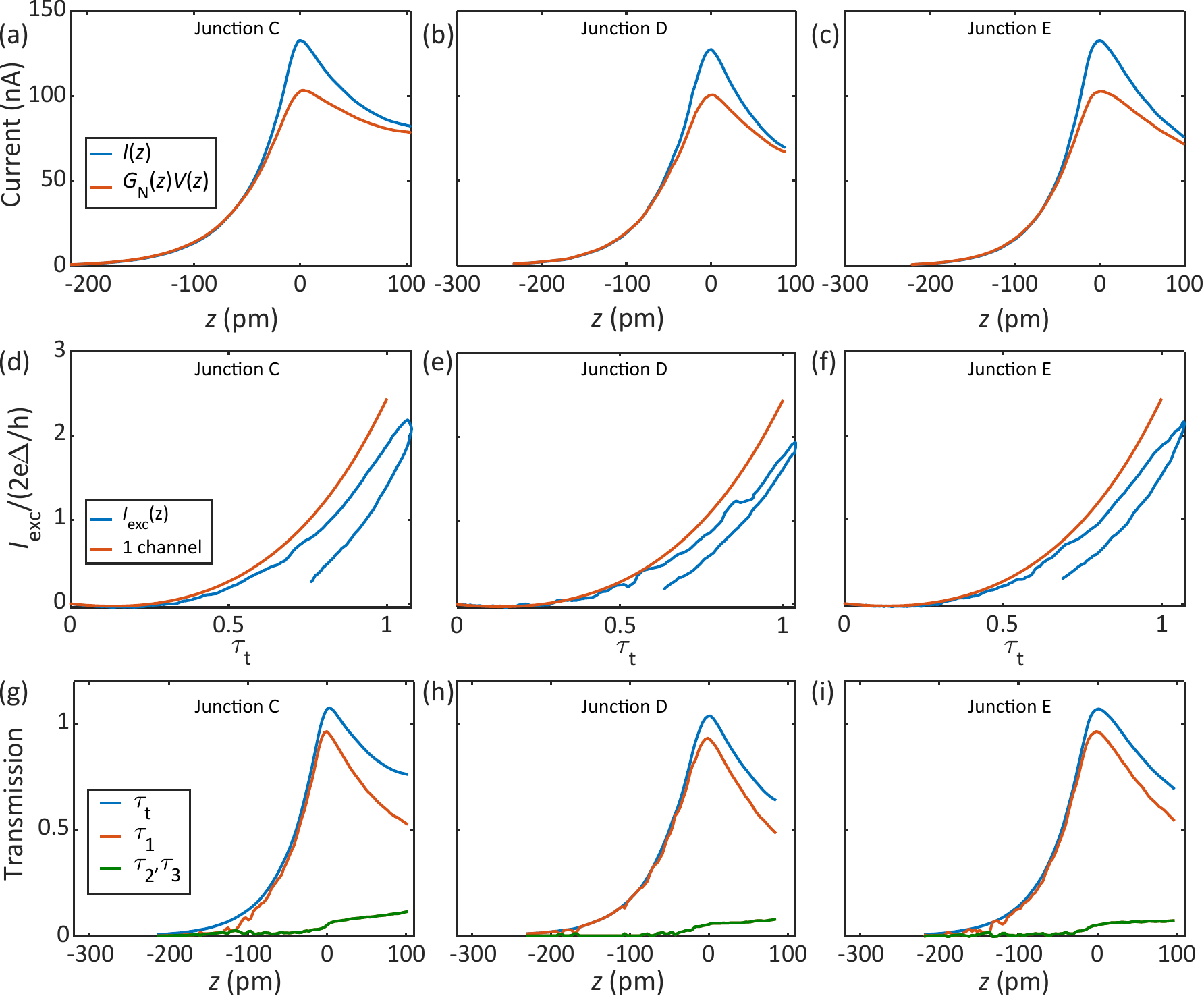}
	\caption{PIN code analysis based on the $I_\text{Exc}$ method.
	In (a) the $I(z)$ curve of junction C is compared to its ohmic equivalent, yielding the excess current $I_\text{Exc}$, displayed in (d). In (g) the extracted PIN code is shown in as function of $z$.	In (b,e,h) and (c,f,i) the same is shown for junctions D and E, respectively.}
	\label{fig:SIzPIN}
\end{figure}

We applied the excess current method developed in this manuscript to further single atom contacts. The results from three additional junctions, which are not shown in the main text, are displayed in \figref{fig:SIzPIN}. All three contacts show a qualitatively similar behaviour to junctions A and B of the main text. The conductance is dominated by a single transport channel. The transmission through the dominant channel peaks at the maximum total transmission and decreased thereafter. The secondary transport channels are negligible up until the maximum conductance is reached and only gain importance upon further approach to the surface.

\begin{figure}[b]
	\centering
	\includegraphics[width=\linewidth]{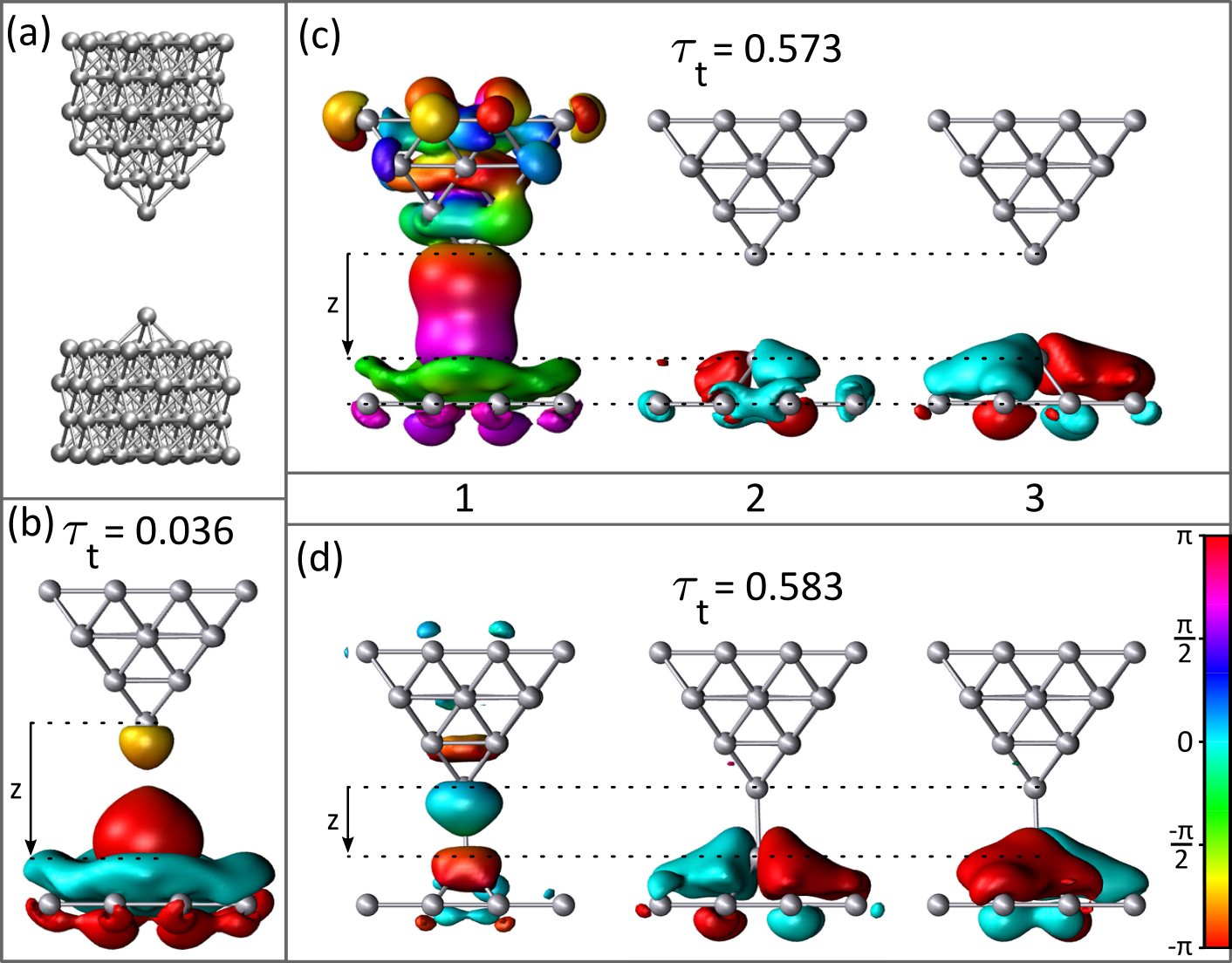}
	\caption{Transport channels obtained from the DFT+NEGF calculations. (a) Tip and sample structure used for the simulations. (b) Wave function of the dominant channel in the tunnelling limit. There is little overlap between the tip and sample. (c) From left to right, wave functions corresponding to the dominant and the two degenerate channels at total contact transparency of 0.573 before the maximum of conductance. The wave function of the dominant channel can no longer be clearly assigned to the tip or sample. The degenerate channels are localised on the sample an do not significantly extend into the vacuum gap. (c) From left to right, wave functions corresponding to the dominant and the two degenerate channels at total contact transparency of 0.583 after the maximum of conductance. The dominant channel resembles a bound state in the junction, while the degenerate channels begin to bridge the gap.}
	\label{figSI:junctionorbs}
\end{figure}

\section{DFT+NEGF results}

We simulated single atom junctions akin to those in the experiment using a combination of density functional theory (DFT) and non-equilibrium Green's function (NEGF) techniques. These calculations yield the optimized geometries and the electronic structure of the junctions as well as forces \cite{si_Pauly2008, si_Buerkle2013}. We model the experimental situation by an Al(100) slab with an Al adatom and tip structures with (100) and (111) orientation. These tip geometries are the closest feasible assumptions for the real, likely amorphous, structure of the tips.

The electronic channels in the contact can be represented by wave functions. For the (100)-oriented tip they are shown in Fig.\ \ref{figSI:junctionorbs} at different tip-sample separations. \figref{figSI:junctionorbs}(a) shows the atomic structure used in the DFT+NEGF calculations. The wave function of the dominant transmission channel in the tunneling limit $\tau_\text{t} \ll 1$ is shown in \figref{figSI:junctionorbs}(b). There is little overlap between the tip and sample electrodes in this limit. The situation changes as tip and sample approach each other and the conductance increases. \figref{figSI:junctionorbs}(c) shows the same junction with $\tau_t \approx 0.58\,G_0$ before the maximum of conductance is reached ($z < 0$). The wave functions of the dominant and degenerate channels are plotted from left to right. The dominant transport channel extends across the tip and sample electrodes, enabling the efficient transfer of charges from one to the other. The situation changes again after reaching the point of maximum conductance at $z=0$. \figref{figSI:junctionorbs}(c) shows the same wave functions as panel (b) at a similar total conductance, but this time past the point of maximum conductance ($z > 0$). The dominant transport channel has changed character and now resembles a bound state within the contact. The degenerate channels, on the other hand, begin to bridge the vacuum gap and thus become more relevant for charge transport. We offer a qualitative understanding of this behaviour in a one-dimensional toy model, see below.


\begin{thebibliography}{38}%
\makeatletter
\providecommand \@ifxundefined [1]{%
 \@ifx{#1\undefined}
}%
\providecommand \@ifnum [1]{%
 \ifnum #1\expandafter \@firstoftwo
 \else \expandafter \@secondoftwo
 \fi
}%
\providecommand \@ifx [1]{%
 \ifx #1\expandafter \@firstoftwo
 \else \expandafter \@secondoftwo
 \fi
}%
\providecommand \natexlab [1]{#1}%
\providecommand \enquote  [1]{``#1''}%
\providecommand \bibnamefont  [1]{#1}%
\providecommand \bibfnamefont [1]{#1}%
\providecommand \citenamefont [1]{#1}%
\providecommand \href@noop [0]{\@secondoftwo}%
\providecommand \href [0]{\begingroup \@sanitize@url \@href}%
\providecommand \@href[1]{\@@startlink{#1}\@@href}%
\providecommand \@@href[1]{\endgroup#1\@@endlink}%
\providecommand \@sanitize@url [0]{\catcode `\\12\catcode `\$12\catcode
  `\&12\catcode `\#12\catcode `\^12\catcode `\_12\catcode `\%12\relax}%
\providecommand \@@startlink[1]{}%
\providecommand \@@endlink[0]{}%
\providecommand \url  [0]{\begingroup\@sanitize@url \@url }%
\providecommand \@url [1]{\endgroup\@href {#1}{\urlprefix }}%
\providecommand \urlprefix  [0]{URL }%
\providecommand \Eprint [0]{\href }%
\providecommand \doibase [0]{https://doi.org/}%
\providecommand \selectlanguage [0]{\@gobble}%
\providecommand \bibinfo  [0]{\@secondoftwo}%
\providecommand \bibfield  [0]{\@secondoftwo}%
\providecommand \translation [1]{[#1]}%
\providecommand \BibitemOpen [0]{}%
\providecommand \bibitemStop [0]{}%
\providecommand \bibitemNoStop [0]{.\EOS\space}%
\providecommand \EOS [0]{\spacefactor3000\relax}%
\providecommand \BibitemShut  [1]{\csname bibitem#1\endcsname}%
\let\auto@bib@innerbib\@empty
\bibitem [{\citenamefont {Landauer}(1957)}]{landauer1957spatial}%
  \BibitemOpen
  \bibfield  {author} {\bibinfo {author} {\bibfnamefont {R.}~\bibnamefont
  {Landauer}},\ }\bibfield  {title} {\bibinfo {title} {Spatial variation of
  currents and fields due to localized scatterers in metallic conduction},\
  }\href@noop {} {\bibfield  {journal} {\bibinfo  {journal} {IBM Journal of
  research and development}\ }\textbf {\bibinfo {volume} {1}},\ \bibinfo
  {pages} {223} (\bibinfo {year} {1957})}\BibitemShut {NoStop}%
\bibitem [{\citenamefont {Landauer}(1970)}]{Landauer1970}%
  \BibitemOpen
  \bibfield  {author} {\bibinfo {author} {\bibfnamefont {R.}~\bibnamefont
  {Landauer}},\ }\bibfield  {title} {\bibinfo {title} {Electrical resistance of
  disordered one-dimensional lattices},\ }\href
  {https://doi.org/10.1080/14786437008238472} {\bibfield  {journal} {\bibinfo
  {journal} {Philos. Mag.}\ }\textbf {\bibinfo {volume} {21}},\ \bibinfo
  {pages} {863} (\bibinfo {year} {1970})}\BibitemShut {NoStop}%
\bibitem [{\citenamefont {Cuevas}(1999)}]{Cuevas1999}%
  \BibitemOpen
  \bibfield  {author} {\bibinfo {author} {\bibfnamefont {J.~C.}\ \bibnamefont
  {Cuevas}},\ }\emph {\bibinfo {title} {Electronic transport in normal and
  superconducting nanostructures}},\ \href
  {http://www.uam.es/personal_pdi/ciencias/jcuevas/Publications/thesis.pdf}
  {Ph.D. thesis},\ \bibinfo  {school} {Universidad Aut\'onoma de Madrid}
  (\bibinfo {year} {1999})\BibitemShut {NoStop}%
\bibitem [{\citenamefont {van~den Brom}\ and\ \citenamefont {van
  Ruitenbeek}(1999)}]{Brom1999}%
  \BibitemOpen
  \bibfield  {author} {\bibinfo {author} {\bibfnamefont {H.~E.}\ \bibnamefont
  {van~den Brom}}\ and\ \bibinfo {author} {\bibfnamefont {J.~M.}\ \bibnamefont
  {van Ruitenbeek}},\ }\bibfield  {title} {\bibinfo {title} {Quantum
  {Suppression} of {Shot Noise in Atom-Size Metallic Contacts}},\ }\href
  {https://doi.org/10.1103/physrevlett.82.1526} {\bibfield  {journal} {\bibinfo
   {journal} {Phys. Rev. Lett.}\ }\textbf {\bibinfo {volume} {82}},\ \bibinfo
  {pages} {1526} (\bibinfo {year} {1999})}\BibitemShut {NoStop}%
\bibitem [{\citenamefont {Schoelkopf}\ \emph {et~al.}(1997)\citenamefont
  {Schoelkopf}, \citenamefont {Burke}, \citenamefont {Kozhevnikov},
  \citenamefont {Prober},\ and\ \citenamefont {Rooks}}]{schoelkopf1997}%
  \BibitemOpen
  \bibfield  {author} {\bibinfo {author} {\bibfnamefont {R.~J.}\ \bibnamefont
  {Schoelkopf}}, \bibinfo {author} {\bibfnamefont {P.~J.}\ \bibnamefont
  {Burke}}, \bibinfo {author} {\bibfnamefont {A.~A.}\ \bibnamefont
  {Kozhevnikov}}, \bibinfo {author} {\bibfnamefont {D.~E.}\ \bibnamefont
  {Prober}},\ and\ \bibinfo {author} {\bibfnamefont {M.~J.}\ \bibnamefont
  {Rooks}},\ }\bibfield  {title} {\bibinfo {title} {{Frequency Dependence of
  Shot Noise in a Diffusive Mesoscopic Conductor}},\ }\href
  {https://doi.org/10.1103/PhysRevLett.78.3370} {\bibfield  {journal} {\bibinfo
   {journal} {Phys. Rev. Lett.}\ }\textbf {\bibinfo {volume} {78}},\ \bibinfo
  {pages} {3370} (\bibinfo {year} {1997})}\BibitemShut {NoStop}%
\bibitem [{\citenamefont {Blanter}\ and\ \citenamefont
  {B\"{u}ttiker}(2000)}]{Blanter2000}%
  \BibitemOpen
  \bibfield  {author} {\bibinfo {author} {\bibfnamefont {Y.}~\bibnamefont
  {Blanter}}\ and\ \bibinfo {author} {\bibfnamefont {M.}~\bibnamefont
  {B\"{u}ttiker}},\ }\bibfield  {title} {\bibinfo {title} {Shot noise in
  mesoscopic conductors},\ }\href
  {https://doi.org/10.1016/s0370-1573(99)00123-4} {\bibfield  {journal}
  {\bibinfo  {journal} {Phys. Rep.}\ }\textbf {\bibinfo {volume} {336}},\
  \bibinfo {pages} {1} (\bibinfo {year} {2000})}\BibitemShut {NoStop}%
\bibitem [{\citenamefont {Cron}\ \emph {et~al.}(2001)\citenamefont {Cron},
  \citenamefont {Goffman}, \citenamefont {Esteve},\ and\ \citenamefont
  {Urbina}}]{Cron2001a}%
  \BibitemOpen
  \bibfield  {author} {\bibinfo {author} {\bibfnamefont {R.}~\bibnamefont
  {Cron}}, \bibinfo {author} {\bibfnamefont {M.~F.}\ \bibnamefont {Goffman}},
  \bibinfo {author} {\bibfnamefont {D.}~\bibnamefont {Esteve}},\ and\ \bibinfo
  {author} {\bibfnamefont {C.}~\bibnamefont {Urbina}},\ }\bibfield  {title}
  {\bibinfo {title} {{Multiple-Charge-Quanta Shot Noise in Superconducting
  Atomic Contacts}},\ }\href {https://doi.org/10.1103/physrevlett.86.4104}
  {\bibfield  {journal} {\bibinfo  {journal} {Phys. Rev. Lett.}\ }\textbf
  {\bibinfo {volume} {86}},\ \bibinfo {pages} {4104} (\bibinfo {year}
  {2001})}\BibitemShut {NoStop}%
\bibitem [{\citenamefont {Agra\"it}(2003)}]{Agrait2003}%
  \BibitemOpen
  \bibfield  {author} {\bibinfo {author} {\bibfnamefont {N.}~\bibnamefont
  {Agra\"it}},\ }\bibfield  {title} {\bibinfo {title} {Quantum properties of
  atomic-sized conductors},\ }\href
  {https://doi.org/10.1016/s0370-1573(02)00633-6} {\bibfield  {journal}
  {\bibinfo  {journal} {Phys. Rep.}\ }\textbf {\bibinfo {volume} {377}},\
  \bibinfo {pages} {81} (\bibinfo {year} {2003})}\BibitemShut {NoStop}%
\bibitem [{\citenamefont {Djukic}\ and\ \citenamefont {van
  Ruitenbeek}(2006)}]{Djukic2006}%
  \BibitemOpen
  \bibfield  {author} {\bibinfo {author} {\bibfnamefont {D.}~\bibnamefont
  {Djukic}}\ and\ \bibinfo {author} {\bibfnamefont {J.~M.}\ \bibnamefont {van
  Ruitenbeek}},\ }\bibfield  {title} {\bibinfo {title} {Shot noise measurements
  on a single molecule},\ }\href {https://doi.org/10.1021/nl060116e} {\bibfield
   {journal} {\bibinfo  {journal} {Nano Lett.}\ }\textbf {\bibinfo {volume}
  {6}},\ \bibinfo {pages} {789} (\bibinfo {year} {2006})}\BibitemShut {NoStop}%
\bibitem [{\citenamefont {Kumar}\ \emph {et~al.}(2013)\citenamefont {Kumar},
  \citenamefont {Tal}, \citenamefont {Smit}, \citenamefont {Smogunov},
  \citenamefont {Tosatti},\ and\ \citenamefont {van Ruitenbeek}}]{Kumar2013}%
  \BibitemOpen
  \bibfield  {author} {\bibinfo {author} {\bibfnamefont {M.}~\bibnamefont
  {Kumar}}, \bibinfo {author} {\bibfnamefont {O.}~\bibnamefont {Tal}}, \bibinfo
  {author} {\bibfnamefont {R.~H.~M.}\ \bibnamefont {Smit}}, \bibinfo {author}
  {\bibfnamefont {A.}~\bibnamefont {Smogunov}}, \bibinfo {author}
  {\bibfnamefont {E.}~\bibnamefont {Tosatti}},\ and\ \bibinfo {author}
  {\bibfnamefont {J.~M.}\ \bibnamefont {van Ruitenbeek}},\ }\bibfield  {title}
  {\bibinfo {title} {Shot noise and magnetism of {Pt} atomic chains:
  Accumulation of points at the boundary},\ }\href
  {https://doi.org/10.1103/physrevb.88.245431} {\bibfield  {journal} {\bibinfo
  {journal} {Phys. Rev. B}\ }\textbf {\bibinfo {volume} {88}},\ \bibinfo
  {pages} {245431} (\bibinfo {year} {2013})}\BibitemShut {NoStop}%
\bibitem [{\citenamefont {Vardimon}\ \emph {et~al.}(2013)\citenamefont
  {Vardimon}, \citenamefont {Klionsky},\ and\ \citenamefont
  {Tal}}]{vardimon2013}%
  \BibitemOpen
  \bibfield  {author} {\bibinfo {author} {\bibfnamefont {R.}~\bibnamefont
  {Vardimon}}, \bibinfo {author} {\bibfnamefont {M.}~\bibnamefont {Klionsky}},\
  and\ \bibinfo {author} {\bibfnamefont {O.}~\bibnamefont {Tal}},\ }\bibfield
  {title} {\bibinfo {title} {Experimental determination of conduction channels
  in atomic-scale conductors based on shot noise measurements},\ }\href
  {https://doi.org/10.1103/physrevb.88.161404} {\bibfield  {journal} {\bibinfo
  {journal} {Phys. Rev. B}\ }\textbf {\bibinfo {volume} {88}},\ \bibinfo
  {pages} {161404} (\bibinfo {year} {2013})}\BibitemShut {NoStop}%
\bibitem [{\citenamefont {Burtzlaff}\ \emph {et~al.}(2015)\citenamefont
  {Burtzlaff}, \citenamefont {Weismann}, \citenamefont {Brandbyge},\ and\
  \citenamefont {Berndt}}]{burtzlaff2015shot}%
  \BibitemOpen
  \bibfield  {author} {\bibinfo {author} {\bibfnamefont {A.}~\bibnamefont
  {Burtzlaff}}, \bibinfo {author} {\bibfnamefont {A.}~\bibnamefont {Weismann}},
  \bibinfo {author} {\bibfnamefont {M.}~\bibnamefont {Brandbyge}},\ and\
  \bibinfo {author} {\bibfnamefont {R.}~\bibnamefont {Berndt}},\ }\bibfield
  {title} {\bibinfo {title} {{Shot Noise as a Probe of Spin-Polarized Transport
  through Single Atoms}},\ }\href
  {https://doi.org/10.1103/physrevlett.114.016602} {\bibfield  {journal}
  {\bibinfo  {journal} {Phys. Rev. Lett.}\ }\textbf {\bibinfo {volume} {114}},\
  \bibinfo {pages} {016602} (\bibinfo {year} {2015})}\BibitemShut {NoStop}%
\bibitem [{\citenamefont {Vardimon}\ \emph {et~al.}(2015)\citenamefont
  {Vardimon}, \citenamefont {Klionsky},\ and\ \citenamefont
  {Tal}}]{Vardimon2015}%
  \BibitemOpen
  \bibfield  {author} {\bibinfo {author} {\bibfnamefont {R.}~\bibnamefont
  {Vardimon}}, \bibinfo {author} {\bibfnamefont {M.}~\bibnamefont {Klionsky}},\
  and\ \bibinfo {author} {\bibfnamefont {O.}~\bibnamefont {Tal}},\ }\bibfield
  {title} {\bibinfo {title} {Indication of complete spin filtering in
  atomic-scale nickel oxide},\ }\href
  {https://doi.org/10.1021/acs.nanolett.5b00729} {\bibfield  {journal}
  {\bibinfo  {journal} {Nano Lett.}\ }\textbf {\bibinfo {volume} {15}},\
  \bibinfo {pages} {3894} (\bibinfo {year} {2015})}\BibitemShut {NoStop}%
\bibitem [{\citenamefont {Vardimon}\ \emph {et~al.}(2016)\citenamefont
  {Vardimon}, \citenamefont {Matt}, \citenamefont {Nielaba}, \citenamefont
  {Cuevas},\ and\ \citenamefont {Tal}}]{Vardimon2016}%
  \BibitemOpen
  \bibfield  {author} {\bibinfo {author} {\bibfnamefont {R.}~\bibnamefont
  {Vardimon}}, \bibinfo {author} {\bibfnamefont {M.}~\bibnamefont {Matt}},
  \bibinfo {author} {\bibfnamefont {P.}~\bibnamefont {Nielaba}}, \bibinfo
  {author} {\bibfnamefont {J.~C.}\ \bibnamefont {Cuevas}},\ and\ \bibinfo
  {author} {\bibfnamefont {O.}~\bibnamefont {Tal}},\ }\bibfield  {title}
  {\bibinfo {title} {Orbital origin of the electrical conduction in
  ferromagnetic atomic-size contacts: Insights from shot noise measurements and
  theoretical simulations},\ }\href
  {https://doi.org/10.1103/physrevb.93.085439} {\bibfield  {journal} {\bibinfo
  {journal} {Phys. Rev. B}\ }\textbf {\bibinfo {volume} {93}},\ \bibinfo
  {pages} {085439} (\bibinfo {year} {2016})}\BibitemShut {NoStop}%
\bibitem [{\citenamefont {Senkpiel}\ \emph
  {et~al.}(2020{\natexlab{a}})\citenamefont {Senkpiel}, \citenamefont
  {Kl{\"o}ckner}, \citenamefont {Etzkorn}, \citenamefont {Dambach},
  \citenamefont {Kubala}, \citenamefont {Belzig}, \citenamefont {Yeyati},
  \citenamefont {Cuevas}, \citenamefont {Pauly}, \citenamefont {Ankerhold}
  \emph {et~al.}}]{senkpiel2020dynamical}%
  \BibitemOpen
  \bibfield  {author} {\bibinfo {author} {\bibfnamefont {J.}~\bibnamefont
  {Senkpiel}}, \bibinfo {author} {\bibfnamefont {J.~C.}\ \bibnamefont
  {Kl{\"o}ckner}}, \bibinfo {author} {\bibfnamefont {M.}~\bibnamefont
  {Etzkorn}}, \bibinfo {author} {\bibfnamefont {S.}~\bibnamefont {Dambach}},
  \bibinfo {author} {\bibfnamefont {B.}~\bibnamefont {Kubala}}, \bibinfo
  {author} {\bibfnamefont {W.}~\bibnamefont {Belzig}}, \bibinfo {author}
  {\bibfnamefont {A.~L.}\ \bibnamefont {Yeyati}}, \bibinfo {author}
  {\bibfnamefont {J.~C.}\ \bibnamefont {Cuevas}}, \bibinfo {author}
  {\bibfnamefont {F.}~\bibnamefont {Pauly}}, \bibinfo {author} {\bibfnamefont
  {J.}~\bibnamefont {Ankerhold}}, \emph {et~al.},\ }\bibfield  {title}
  {\bibinfo {title} {{Dynamical Coulomb Blockade as a Local Probe for Quantum
  Transport}},\ }\href {https://doi.org/10.1103/PhysRevLett.124.156803}
  {\bibfield  {journal} {\bibinfo  {journal} {Phys. Rev. Lett.}\ }\textbf
  {\bibinfo {volume} {124}},\ \bibinfo {pages} {156803} (\bibinfo {year}
  {2020}{\natexlab{a}})}\BibitemShut {NoStop}%
\bibitem [{\citenamefont {Muller}\ \emph {et~al.}(1992)\citenamefont {Muller},
  \citenamefont {van Ruitenbeek},\ and\ \citenamefont {de~Jongh}}]{Muller1992}%
  \BibitemOpen
  \bibfield  {author} {\bibinfo {author} {\bibfnamefont {C.~J.}\ \bibnamefont
  {Muller}}, \bibinfo {author} {\bibfnamefont {J.~M.}\ \bibnamefont {van
  Ruitenbeek}},\ and\ \bibinfo {author} {\bibfnamefont {L.~J.}\ \bibnamefont
  {de~Jongh}},\ }\bibfield  {title} {\bibinfo {title} {Experimental observation
  of the transition from weak link to tunnel junction},\ }\href
  {https://doi.org/10.1016/0921-4534(92)90947-b} {\bibfield  {journal}
  {\bibinfo  {journal} {Physica C}\ }\textbf {\bibinfo {volume} {191}},\
  \bibinfo {pages} {485} (\bibinfo {year} {1992})}\BibitemShut {NoStop}%
\bibitem [{\citenamefont {Scheer}\ \emph {et~al.}(1997)\citenamefont {Scheer},
  \citenamefont {Joyez}, \citenamefont {Esteve}, \citenamefont {Urbina},\ and\
  \citenamefont {Devoret}}]{Scheer1997}%
  \BibitemOpen
  \bibfield  {author} {\bibinfo {author} {\bibfnamefont {E.}~\bibnamefont
  {Scheer}}, \bibinfo {author} {\bibfnamefont {P.}~\bibnamefont {Joyez}},
  \bibinfo {author} {\bibfnamefont {D.}~\bibnamefont {Esteve}}, \bibinfo
  {author} {\bibfnamefont {C.}~\bibnamefont {Urbina}},\ and\ \bibinfo {author}
  {\bibfnamefont {M.~H.}\ \bibnamefont {Devoret}},\ }\bibfield  {title}
  {\bibinfo {title} {{Conduction Channel Transmissions of Atomic-Size Aluminum
  Contacts}},\ }\href {https://doi.org/10.1103/physrevlett.78.3535} {\bibfield
  {journal} {\bibinfo  {journal} {Phys. Rev. Lett.}\ }\textbf {\bibinfo
  {volume} {78}},\ \bibinfo {pages} {3535} (\bibinfo {year}
  {1997})}\BibitemShut {NoStop}%
\bibitem [{\citenamefont {Scheer}\ \emph {et~al.}(1998)\citenamefont {Scheer},
  \citenamefont {Agra{\"{i}}t}, \citenamefont {Cuevas}, \citenamefont {{Levy
  Yeyati}}, \citenamefont {Ludoph}, \citenamefont {Mart{\'{\i}}n-Rodero},
  \citenamefont {Bollinger}, \citenamefont {van Ruitenbeek},\ and\
  \citenamefont {Urbina}}]{Scheer1998}%
  \BibitemOpen
  \bibfield  {author} {\bibinfo {author} {\bibfnamefont {E.}~\bibnamefont
  {Scheer}}, \bibinfo {author} {\bibfnamefont {N.}~\bibnamefont
  {Agra{\"{i}}t}}, \bibinfo {author} {\bibfnamefont {J.~C.}\ \bibnamefont
  {Cuevas}}, \bibinfo {author} {\bibfnamefont {A.}~\bibnamefont {{Levy
  Yeyati}}}, \bibinfo {author} {\bibfnamefont {B.}~\bibnamefont {Ludoph}},
  \bibinfo {author} {\bibfnamefont {A.}~\bibnamefont {Mart{\'{\i}}n-Rodero}},
  \bibinfo {author} {\bibfnamefont {G.~R.}\ \bibnamefont {Bollinger}}, \bibinfo
  {author} {\bibfnamefont {J.~M.}\ \bibnamefont {van Ruitenbeek}},\ and\
  \bibinfo {author} {\bibfnamefont {C.}~\bibnamefont {Urbina}},\ }\bibfield
  {title} {\bibinfo {title} {The signature of chemical valence in the
  electrical conduction through a single-atom contact},\ }\href
  {https://doi.org/10.1038/28112} {\bibfield  {journal} {\bibinfo  {journal}
  {Nature}\ }\textbf {\bibinfo {volume} {394}},\ \bibinfo {pages} {154}
  (\bibinfo {year} {1998})}\BibitemShut {NoStop}%
\bibitem [{\citenamefont {Chauvin}\ \emph {et~al.}(2007)\citenamefont
  {Chauvin}, \citenamefont {vom Stein}, \citenamefont {Esteve}, \citenamefont
  {Urbina}, \citenamefont {Cuevas},\ and\ \citenamefont {{Levy
  Yeyati}}}]{Chauvin2007}%
  \BibitemOpen
  \bibfield  {author} {\bibinfo {author} {\bibfnamefont {M.}~\bibnamefont
  {Chauvin}}, \bibinfo {author} {\bibfnamefont {P.}~\bibnamefont {vom Stein}},
  \bibinfo {author} {\bibfnamefont {D.}~\bibnamefont {Esteve}}, \bibinfo
  {author} {\bibfnamefont {C.}~\bibnamefont {Urbina}}, \bibinfo {author}
  {\bibfnamefont {J.~C.}\ \bibnamefont {Cuevas}},\ and\ \bibinfo {author}
  {\bibfnamefont {A.}~\bibnamefont {{Levy Yeyati}}},\ }\bibfield  {title}
  {\bibinfo {title} {{Crossover from {J}osephson to Multiple {A}ndreev
  Reflection Currents in Atomic Contacts}},\ }\href
  {https://doi.org/10.1103/physrevlett.99.067008} {\bibfield  {journal}
  {\bibinfo  {journal} {Phys. Rev. Lett.}\ }\textbf {\bibinfo {volume} {99}},\
  \bibinfo {pages} {067008} (\bibinfo {year} {2007})}\BibitemShut {NoStop}%
\bibitem [{\citenamefont {Rocca}\ \emph {et~al.}(2007)\citenamefont {Rocca},
  \citenamefont {Chauvin}, \citenamefont {Huard}, \citenamefont {Pothier},
  \citenamefont {Esteve},\ and\ \citenamefont {Urbina}}]{Rocca2007}%
  \BibitemOpen
  \bibfield  {author} {\bibinfo {author} {\bibfnamefont {M.~L.~D.}\
  \bibnamefont {Rocca}}, \bibinfo {author} {\bibfnamefont {M.}~\bibnamefont
  {Chauvin}}, \bibinfo {author} {\bibfnamefont {B.}~\bibnamefont {Huard}},
  \bibinfo {author} {\bibfnamefont {H.}~\bibnamefont {Pothier}}, \bibinfo
  {author} {\bibfnamefont {D.}~\bibnamefont {Esteve}},\ and\ \bibinfo {author}
  {\bibfnamefont {C.}~\bibnamefont {Urbina}},\ }\bibfield  {title} {\bibinfo
  {title} {{Measurement of the Current-Phase Relation of Superconducting Atomic
  Contacts}},\ }\href {https://doi.org/10.1103/physrevlett.99.127005}
  {\bibfield  {journal} {\bibinfo  {journal} {Phys. Rev. Lett.}\ }\textbf
  {\bibinfo {volume} {99}},\ \bibinfo {pages} {127005} (\bibinfo {year}
  {2007})}\BibitemShut {NoStop}%
\bibitem [{\citenamefont {Ludoph}\ \emph {et~al.}(2000)\citenamefont {Ludoph},
  \citenamefont {van~der Post}, \citenamefont {Bratus}, \citenamefont
  {Bezuglyi}, \citenamefont {Shumeiko}, \citenamefont {Wendin},\ and\
  \citenamefont {van Ruitenbeek}}]{ludoph2000multiple}%
  \BibitemOpen
  \bibfield  {author} {\bibinfo {author} {\bibfnamefont {B.}~\bibnamefont
  {Ludoph}}, \bibinfo {author} {\bibfnamefont {N.}~\bibnamefont {van~der
  Post}}, \bibinfo {author} {\bibfnamefont {E.~N.}\ \bibnamefont {Bratus}},
  \bibinfo {author} {\bibfnamefont {E.~V.}\ \bibnamefont {Bezuglyi}}, \bibinfo
  {author} {\bibfnamefont {V.~S.}\ \bibnamefont {Shumeiko}}, \bibinfo {author}
  {\bibfnamefont {G.}~\bibnamefont {Wendin}},\ and\ \bibinfo {author}
  {\bibfnamefont {J.~M.}\ \bibnamefont {van Ruitenbeek}},\ }\bibfield  {title}
  {\bibinfo {title} {{Multiple Andreev reflection in single-atom niobium
  junctions}},\ }\href {https://doi.org/10.1103/PhysRevB.61.8561} {\bibfield
  {journal} {\bibinfo  {journal} {Physical Review B}\ }\textbf {\bibinfo
  {volume} {61}},\ \bibinfo {pages} {8561} (\bibinfo {year}
  {2000})}\BibitemShut {NoStop}%
\bibitem [{\citenamefont {Riquelme}\ \emph {et~al.}(2005)\citenamefont
  {Riquelme}, \citenamefont {de~la Vega}, \citenamefont {Yeyati}, \citenamefont
  {Agra{\"\i}t}, \citenamefont {Martin-Rodero},\ and\ \citenamefont
  {Rubio-Bollinger}}]{riquelme2005distribution}%
  \BibitemOpen
  \bibfield  {author} {\bibinfo {author} {\bibfnamefont {J.}~\bibnamefont
  {Riquelme}}, \bibinfo {author} {\bibfnamefont {L.}~\bibnamefont {de~la
  Vega}}, \bibinfo {author} {\bibfnamefont {A.~L.}\ \bibnamefont {Yeyati}},
  \bibinfo {author} {\bibfnamefont {N.}~\bibnamefont {Agra{\"\i}t}}, \bibinfo
  {author} {\bibfnamefont {A.}~\bibnamefont {Martin-Rodero}},\ and\ \bibinfo
  {author} {\bibfnamefont {G.}~\bibnamefont {Rubio-Bollinger}},\ }\bibfield
  {title} {\bibinfo {title} {Distribution of conduction channels in nanoscale
  contacts: Evolution towards the diffusive limit},\ }\href
  {https://doi.org/10.1209/epl/i2005-10028-0} {\bibfield  {journal} {\bibinfo
  {journal} {Europhys. Lett.}\ }\textbf {\bibinfo {volume} {70}},\ \bibinfo
  {pages} {663} (\bibinfo {year} {2005})}\BibitemShut {NoStop}%
\bibitem [{\citenamefont {Massee}\ \emph {et~al.}(2018)\citenamefont {Massee},
  \citenamefont {Dong}, \citenamefont {Cavanna}, \citenamefont {Jin},\ and\
  \citenamefont {Aprili}}]{Massee2018}%
  \BibitemOpen
  \bibfield  {author} {\bibinfo {author} {\bibfnamefont {F.}~\bibnamefont
  {Massee}}, \bibinfo {author} {\bibfnamefont {Q.}~\bibnamefont {Dong}},
  \bibinfo {author} {\bibfnamefont {A.}~\bibnamefont {Cavanna}}, \bibinfo
  {author} {\bibfnamefont {Y.}~\bibnamefont {Jin}},\ and\ \bibinfo {author}
  {\bibfnamefont {M.}~\bibnamefont {Aprili}},\ }\bibfield  {title} {\bibinfo
  {title} {Atomic scale shot-noise using cryogenic {MHz} circuitry},\ }\href
  {https://doi.org/10.1063/1.5043261} {\bibfield  {journal} {\bibinfo
  {journal} {Rev. Sci. Instrum.}\ }\textbf {\bibinfo {volume} {89}},\ \bibinfo
  {pages} {093708} (\bibinfo {year} {2018})}\BibitemShut {NoStop}%
\bibitem [{\citenamefont {Massee}\ \emph {et~al.}(2019)\citenamefont {Massee},
  \citenamefont {Huang}, \citenamefont {Golden},\ and\ \citenamefont
  {Aprili}}]{Massee2019}%
  \BibitemOpen
  \bibfield  {author} {\bibinfo {author} {\bibfnamefont {F.}~\bibnamefont
  {Massee}}, \bibinfo {author} {\bibfnamefont {Y.~K.}\ \bibnamefont {Huang}},
  \bibinfo {author} {\bibfnamefont {M.~S.}\ \bibnamefont {Golden}},\ and\
  \bibinfo {author} {\bibfnamefont {M.}~\bibnamefont {Aprili}},\ }\bibfield
  {title} {\bibinfo {title} {Noisy defects in the high-${T}_\text{c}$
  superconductor {Bi}$_2${Sr}$_2${CaCu}$_2${O}$_{8+x}$},\ }\href
  {https://doi.org/10.1038/s41467-019-08518-1} {\bibfield  {journal} {\bibinfo
  {journal} {Nat. Commun.}\ }\textbf {\bibinfo {volume} {10}},\ \bibinfo
  {pages} {544} (\bibinfo {year} {2019})}\BibitemShut {NoStop}%
\bibitem [{\citenamefont {Bastiaans}\ \emph {et~al.}(2018)\citenamefont
  {Bastiaans}, \citenamefont {Benschop}, \citenamefont {Chatzopoulos},
  \citenamefont {Cho}, \citenamefont {Dong}, \citenamefont {Jin},\ and\
  \citenamefont {Allan}}]{bastiaans_amplifier_2018}%
  \BibitemOpen
  \bibfield  {author} {\bibinfo {author} {\bibfnamefont {K.~M.}\ \bibnamefont
  {Bastiaans}}, \bibinfo {author} {\bibfnamefont {T.}~\bibnamefont {Benschop}},
  \bibinfo {author} {\bibfnamefont {D.}~\bibnamefont {Chatzopoulos}}, \bibinfo
  {author} {\bibfnamefont {D.}~\bibnamefont {Cho}}, \bibinfo {author}
  {\bibfnamefont {Q.}~\bibnamefont {Dong}}, \bibinfo {author} {\bibfnamefont
  {Y.}~\bibnamefont {Jin}},\ and\ \bibinfo {author} {\bibfnamefont {M.~P.}\
  \bibnamefont {Allan}},\ }\bibfield  {title} {\bibinfo {title} {Amplifier for
  scanning tunneling microscopy at {MHz} frequencies},\ }\href
  {https://doi.org/10.1063/1.5043267} {\bibfield  {journal} {\bibinfo
  {journal} {Rev. Sci. Instrum.}\ }\textbf {\bibinfo {volume} {89}},\ \bibinfo
  {pages} {093709} (\bibinfo {year} {2018})}\BibitemShut {NoStop}%
\bibitem [{\citenamefont {Ludoph}\ \emph {et~al.}(1999)\citenamefont {Ludoph},
  \citenamefont {Devoret}, \citenamefont {Esteve}, \citenamefont {Urbina},\
  and\ \citenamefont {van Ruitenbeek}}]{ludoph1999evidence}%
  \BibitemOpen
  \bibfield  {author} {\bibinfo {author} {\bibfnamefont {B.}~\bibnamefont
  {Ludoph}}, \bibinfo {author} {\bibfnamefont {M.~H.}\ \bibnamefont {Devoret}},
  \bibinfo {author} {\bibfnamefont {D.}~\bibnamefont {Esteve}}, \bibinfo
  {author} {\bibfnamefont {C.}~\bibnamefont {Urbina}},\ and\ \bibinfo {author}
  {\bibfnamefont {J.~M.}\ \bibnamefont {van Ruitenbeek}},\ }\bibfield  {title}
  {\bibinfo {title} {{Evidence for Saturation of Channel Transmission from
  Conductance Fluctuations in Atomic-Size Point Contacts}},\ }\href
  {https://doi.org/10.1103/PhysRevLett.82.1530} {\bibfield  {journal} {\bibinfo
   {journal} {Phys. Rev. Lett.}\ }\textbf {\bibinfo {volume} {82}},\ \bibinfo
  {pages} {1530} (\bibinfo {year} {1999})}\BibitemShut {NoStop}%
\bibitem [{\citenamefont {Ludoph}\ and\ \citenamefont {van
  Ruitenbeek}(2000)}]{ludoph2000conductance}%
  \BibitemOpen
  \bibfield  {author} {\bibinfo {author} {\bibfnamefont {B.}~\bibnamefont
  {Ludoph}}\ and\ \bibinfo {author} {\bibfnamefont {J.~M.}\ \bibnamefont {van
  Ruitenbeek}},\ }\bibfield  {title} {\bibinfo {title} {Conductance
  fluctuations as a tool for investigating the quantum modes in atomic-size
  metallic contacts},\ }\href {https://doi.org/10.1103/PhysRevB.61.2273}
  {\bibfield  {journal} {\bibinfo  {journal} {Phys. Rev. B}\ }\textbf {\bibinfo
  {volume} {61}},\ \bibinfo {pages} {2273} (\bibinfo {year}
  {2000})}\BibitemShut {NoStop}%
\bibitem [{\citenamefont {Andreev}(1964)}]{andreev1964thermal}%
  \BibitemOpen
  \bibfield  {author} {\bibinfo {author} {\bibfnamefont {A.}~\bibnamefont
  {Andreev}},\ }\bibfield  {title} {\bibinfo {title} {The thermal conductivity
  of the intermediate state in superconductors},\ }\href@noop {} {\bibfield
  {journal} {\bibinfo  {journal} {Zh. Eksp. Teor. Fiz.}\ }\textbf {\bibinfo
  {volume} {46}},\ \bibinfo {pages} {1823} (\bibinfo {year}
  {1964})}\BibitemShut {NoStop}%
\bibitem [{\citenamefont {Blonder}\ \emph {et~al.}(1982)\citenamefont
  {Blonder}, \citenamefont {Tinkham},\ and\ \citenamefont
  {Klapwijk}}]{Blonder1982}%
  \BibitemOpen
  \bibfield  {author} {\bibinfo {author} {\bibfnamefont {G.~E.}\ \bibnamefont
  {Blonder}}, \bibinfo {author} {\bibfnamefont {M.}~\bibnamefont {Tinkham}},\
  and\ \bibinfo {author} {\bibfnamefont {T.~M.}\ \bibnamefont {Klapwijk}},\
  }\bibfield  {title} {\bibinfo {title} {Transition from metallic to tunneling
  regimes in superconducting microconstrictions: Excess current, charge
  imbalance, and supercurrent conversion},\ }\href
  {https://doi.org/10.1103/physrevb.25.4515} {\bibfield  {journal} {\bibinfo
  {journal} {Phys. Rev. B}\ }\textbf {\bibinfo {volume} {25}},\ \bibinfo
  {pages} {4515} (\bibinfo {year} {1982})}\BibitemShut {NoStop}%
\bibitem [{\citenamefont {Assig}\ \emph {et~al.}(2013)\citenamefont {Assig},
  \citenamefont {Etzkorn}, \citenamefont {Enders}, \citenamefont {Stiepany},
  \citenamefont {Ast},\ and\ \citenamefont {Kern}}]{Assig2013}%
  \BibitemOpen
  \bibfield  {author} {\bibinfo {author} {\bibfnamefont {M.}~\bibnamefont
  {Assig}}, \bibinfo {author} {\bibfnamefont {M.}~\bibnamefont {Etzkorn}},
  \bibinfo {author} {\bibfnamefont {A.}~\bibnamefont {Enders}}, \bibinfo
  {author} {\bibfnamefont {W.}~\bibnamefont {Stiepany}}, \bibinfo {author}
  {\bibfnamefont {C.~R.}\ \bibnamefont {Ast}},\ and\ \bibinfo {author}
  {\bibfnamefont {K.}~\bibnamefont {Kern}},\ }\bibfield  {title} {\bibinfo
  {title} {A {10\,mK} scanning tunneling microscope operating in ultra high
  vacuum and high magnetic fields},\ }\href {https://doi.org/10.1063/1.4793793}
  {\bibfield  {journal} {\bibinfo  {journal} {Rev. Sci. Instrum.}\ }\textbf
  {\bibinfo {volume} {84}},\ \bibinfo {pages} {033903} (\bibinfo {year}
  {2013})}\BibitemShut {NoStop}%
\bibitem [{si()}]{si}%
  \BibitemOpen
  \href@noop {} {}\bibinfo {note} {See Supplemental Material for additional
  data and further details on sample preparation, absence of jump to contact
  and hysteresis in approach-retract curves, and DFT+NEGF
  simulations.}\BibitemShut {Stop}%
\bibitem [{\citenamefont {Senkpiel}\ \emph
  {et~al.}(2020{\natexlab{b}})\citenamefont {Senkpiel}, \citenamefont
  {Dambach}, \citenamefont {Etzkorn}, \citenamefont {Drost}, \citenamefont
  {Padurariu}, \citenamefont {Kubala}, \citenamefont {Belzig}, \citenamefont
  {Yeyati}, \citenamefont {Cuevas}, \citenamefont {Ankerhold} \emph
  {et~al.}}]{senkpiel2020single}%
  \BibitemOpen
  \bibfield  {author} {\bibinfo {author} {\bibfnamefont {J.}~\bibnamefont
  {Senkpiel}}, \bibinfo {author} {\bibfnamefont {S.}~\bibnamefont {Dambach}},
  \bibinfo {author} {\bibfnamefont {M.}~\bibnamefont {Etzkorn}}, \bibinfo
  {author} {\bibfnamefont {R.}~\bibnamefont {Drost}}, \bibinfo {author}
  {\bibfnamefont {C.}~\bibnamefont {Padurariu}}, \bibinfo {author}
  {\bibfnamefont {B.}~\bibnamefont {Kubala}}, \bibinfo {author} {\bibfnamefont
  {W.}~\bibnamefont {Belzig}}, \bibinfo {author} {\bibfnamefont {A.~L.}\
  \bibnamefont {Yeyati}}, \bibinfo {author} {\bibfnamefont {J.~C.}\
  \bibnamefont {Cuevas}}, \bibinfo {author} {\bibfnamefont {J.}~\bibnamefont
  {Ankerhold}}, \emph {et~al.},\ }\bibfield  {title} {\bibinfo {title} {{Single
  channel Josephson effect in a high transmission atomic contact}},\ }\href
  {https://doi.org/10.1038/s42005-020-00397-z} {\bibfield  {journal} {\bibinfo
  {journal} {Commun. Phys.}\ }\textbf {\bibinfo {volume} {3}},\ \bibinfo
  {pages} {131} (\bibinfo {year} {2020}{\natexlab{b}})}\BibitemShut {NoStop}%
\bibitem [{\citenamefont {Untiedt}\ \emph {et~al.}(2007)\citenamefont
  {Untiedt}, \citenamefont {Caturla}, \citenamefont {Calvo}, \citenamefont
  {Palacios}, \citenamefont {Segers},\ and\ \citenamefont {van
  Ruitenbeek}}]{untiedt2007formation}%
  \BibitemOpen
  \bibfield  {author} {\bibinfo {author} {\bibfnamefont {C.}~\bibnamefont
  {Untiedt}}, \bibinfo {author} {\bibfnamefont {M.~J.}\ \bibnamefont
  {Caturla}}, \bibinfo {author} {\bibfnamefont {M.~R.}\ \bibnamefont {Calvo}},
  \bibinfo {author} {\bibfnamefont {J.~J.}\ \bibnamefont {Palacios}}, \bibinfo
  {author} {\bibfnamefont {R.~C.}\ \bibnamefont {Segers}},\ and\ \bibinfo
  {author} {\bibfnamefont {J.~M.}\ \bibnamefont {van Ruitenbeek}},\ }\bibfield
  {title} {\bibinfo {title} {{Formation} of a {Metallic} {Contact}: {Jump} to
  {Contact} {Revisited}},\ }\href
  {https://doi.org/10.1103/physrevlett.98.206801} {\bibfield  {journal}
  {\bibinfo  {journal} {Phys. Rev. Lett.}\ }\textbf {\bibinfo {volume} {98}},\
  \bibinfo {pages} {206801} (\bibinfo {year} {2007})}\BibitemShut {NoStop}%
\bibitem [{\citenamefont {Cuevas}\ \emph {et~al.}(1996)\citenamefont {Cuevas},
  \citenamefont {Mart{\'{\i}}n-Rodero},\ and\ \citenamefont {{Levy
  Yeyati}}}]{Cuevas1996}%
  \BibitemOpen
  \bibfield  {author} {\bibinfo {author} {\bibfnamefont {J.~C.}\ \bibnamefont
  {Cuevas}}, \bibinfo {author} {\bibfnamefont {A.}~\bibnamefont
  {Mart{\'{\i}}n-Rodero}},\ and\ \bibinfo {author} {\bibfnamefont
  {A.}~\bibnamefont {{Levy Yeyati}}},\ }\bibfield  {title} {\bibinfo {title}
  {Hamiltonian approach to the transport properties of superconducting quantum
  point contacts},\ }\href {https://doi.org/10.1103/physrevb.54.7366}
  {\bibfield  {journal} {\bibinfo  {journal} {Phys. Rev. B}\ }\textbf {\bibinfo
  {volume} {54}},\ \bibinfo {pages} {7366} (\bibinfo {year}
  {1996})}\BibitemShut {NoStop}%
\bibitem [{\citenamefont {Cuevas}\ \emph
  {et~al.}(1998{\natexlab{a}})\citenamefont {Cuevas}, \citenamefont {{Levy
  Yeyati}},\ and\ \citenamefont {Mart{\'{\i}}n-Rodero}}]{Cuevas1998}%
  \BibitemOpen
  \bibfield  {author} {\bibinfo {author} {\bibfnamefont {J.~C.}\ \bibnamefont
  {Cuevas}}, \bibinfo {author} {\bibfnamefont {A.}~\bibnamefont {{Levy
  Yeyati}}},\ and\ \bibinfo {author} {\bibfnamefont {A.}~\bibnamefont
  {Mart{\'{\i}}n-Rodero}},\ }\bibfield  {title} {\bibinfo {title} {{Microscopic
  Origin of Conducting Channels in Metallic Atomic-Size Contacts}},\ }\href
  {https://doi.org/10.1103/physrevlett.80.1066} {\bibfield  {journal} {\bibinfo
   {journal} {Phys. Rev. Lett.}\ }\textbf {\bibinfo {volume} {80}},\ \bibinfo
  {pages} {1066} (\bibinfo {year} {1998}{\natexlab{a}})}\BibitemShut {NoStop}%
\bibitem [{\citenamefont {Dynes}\ \emph {et~al.}(1978)\citenamefont {Dynes},
  \citenamefont {Narayanamurti},\ and\ \citenamefont {Garno}}]{Dynes1978a}%
  \BibitemOpen
  \bibfield  {author} {\bibinfo {author} {\bibfnamefont {R.~C.}\ \bibnamefont
  {Dynes}}, \bibinfo {author} {\bibfnamefont {V.}~\bibnamefont
  {Narayanamurti}},\ and\ \bibinfo {author} {\bibfnamefont {J.~P.}\
  \bibnamefont {Garno}},\ }\bibfield  {title} {\bibinfo {title} {Direct
  measurement of quasiparticle-lifetime broadening in a strong-coupled
  superconductor},\ }\href {https://doi.org/10.1103/physrevlett.41.1509}
  {\bibfield  {journal} {\bibinfo  {journal} {Phys. Rev. Lett.}\ }\textbf
  {\bibinfo {volume} {41}},\ \bibinfo {pages} {1509} (\bibinfo {year}
  {1978})}\BibitemShut {NoStop}%
\bibitem [{\citenamefont {Cuevas}\ \emph
  {et~al.}(1998{\natexlab{b}})\citenamefont {Cuevas}, \citenamefont {{Levy
  Yeyati}}, \citenamefont {Mart{\'{\i}}n-Rodero}, \citenamefont {Bollinger},
  \citenamefont {Untiedt},\ and\ \citenamefont {Agra{\"{i}}t}}]{cuevas1998a}%
  \BibitemOpen
  \bibfield  {author} {\bibinfo {author} {\bibfnamefont {J.~C.}\ \bibnamefont
  {Cuevas}}, \bibinfo {author} {\bibfnamefont {A.}~\bibnamefont {{Levy
  Yeyati}}}, \bibinfo {author} {\bibfnamefont {A.}~\bibnamefont
  {Mart{\'{\i}}n-Rodero}}, \bibinfo {author} {\bibfnamefont {G.~R.}\
  \bibnamefont {Bollinger}}, \bibinfo {author} {\bibfnamefont {C.}~\bibnamefont
  {Untiedt}},\ and\ \bibinfo {author} {\bibfnamefont {N.}~\bibnamefont
  {Agra{\"{i}}t}},\ }\bibfield  {title} {\bibinfo {title} {{Evolution of
  Conducting Channels in Metallic Atomic Contacts under Elastic Deformation}},\
  }\href {https://doi.org/10.1103/physrevlett.81.2990} {\bibfield  {journal}
  {\bibinfo  {journal} {Phys. Rev. Lett.}\ }\textbf {\bibinfo {volume} {81}},\
  \bibinfo {pages} {2990} (\bibinfo {year} {1998}{\natexlab{b}})}\BibitemShut
  {NoStop}%
\bibitem [{\citenamefont {Pauly}\ \emph {et~al.}(2008)\citenamefont {Pauly},
  \citenamefont {Viljas}, \citenamefont {Huniar}, \citenamefont {H\"{a}fner},
  \citenamefont {Wohlthat}, \citenamefont {B\"{u}rkle}, \citenamefont
  {Cuevas},\ and\ \citenamefont {Sch\"{o}n}}]{Pauly2008}%
  \BibitemOpen
  \bibfield  {author} {\bibinfo {author} {\bibfnamefont {F.}~\bibnamefont
  {Pauly}}, \bibinfo {author} {\bibfnamefont {J.~K.}\ \bibnamefont {Viljas}},
  \bibinfo {author} {\bibfnamefont {U.}~\bibnamefont {Huniar}}, \bibinfo
  {author} {\bibfnamefont {M.}~\bibnamefont {H\"{a}fner}}, \bibinfo {author}
  {\bibfnamefont {S.}~\bibnamefont {Wohlthat}}, \bibinfo {author}
  {\bibfnamefont {M.}~\bibnamefont {B\"{u}rkle}}, \bibinfo {author}
  {\bibfnamefont {J.~C.}\ \bibnamefont {Cuevas}},\ and\ \bibinfo {author}
  {\bibfnamefont {G.}~\bibnamefont {Sch\"{o}n}},\ }\bibfield  {title} {\bibinfo
  {title} {Cluster-based density-functional approach to quantum transport
  through molecular and atomic contacts},\ }\href
  {https://doi.org/10.1088/1367-2630/10/12/125019} {\bibfield  {journal}
  {\bibinfo  {journal} {New J. Phys.}\ }\textbf {\bibinfo {volume} {10}},\
  \bibinfo {pages} {125019} (\bibinfo {year} {2008})}\BibitemShut {NoStop}%
\end{thebibliography}

\begin{thebibliography}{2}%
\makeatletter
\providecommand \@ifxundefined [1]{%
 \@ifx{#1\undefined}
}%
\providecommand \@ifnum [1]{%
 \ifnum #1\expandafter \@firstoftwo
 \else \expandafter \@secondoftwo
 \fi
}%
\providecommand \@ifx [1]{%
 \ifx #1\expandafter \@firstoftwo
 \else \expandafter \@secondoftwo
 \fi
}%
\providecommand \natexlab [1]{#1}%
\providecommand \emph  [1]{``#1''}%
\providecommand \bibnamefont  [1]{#1}%
\providecommand \bibfnamefont [1]{#1}%
\providecommand \citenamefont [1]{#1}%
\providecommand \href@noop [0]{\@secondoftwo}%
\providecommand \href [0]{\begingroup \@sanitize@url \@href}%
\providecommand \@href[1]{\@@startlink{#1}\@@href}%
\providecommand \@@href[1]{\endgroup#1\@@endlink}%
\providecommand \@sanitize@url [0]{\catcode `\\12\catcode `\$12\catcode
  `\&12\catcode `\#12\catcode `\^12\catcode `\_12\catcode `\%12\relax}%
\providecommand \@@startlink[1]{}%
\providecommand \@@endlink[0]{}%
\providecommand \url  [0]{\begingroup\@sanitize@url \@url }%
\providecommand \@url [1]{\endgroup\@href {#1}{\urlprefix }}%
\providecommand \urlprefix  [0]{URL }%
\providecommand \Eprint [0]{\href }%
\providecommand \doibase [0]{http://dx.doi.org/}%
\providecommand \selectlanguage [0]{\@gobble}%
\providecommand \bibinfo  [0]{\@secondoftwo}%
\providecommand \bibfield  [0]{\@secondoftwo}%
\providecommand \translation [1]{[#1]}%
\providecommand \BibitemOpen [0]{}%
\providecommand \bibitemStop [0]{}%
\providecommand \bibitemNoStop [0]{.\EOS\space}%
\providecommand \EOS [0]{\spacefactor3000\relax}%
\providecommand \BibitemShut  [1]{\csname bibitem#1\endcsname}%
\let\auto@bib@innerbib\@empty
\bibitem [{\citenamefont {Pauly}\ \emph {et~al.}(2008)\citenamefont {Pauly},
  \citenamefont {Viljas}, \citenamefont {Huniar}, \citenamefont {H\"{a}fner},
  \citenamefont {Wohlthat}, \citenamefont {B\"{u}rkle}, \citenamefont
  {Cuevas},\ and\ \citenamefont {Sch\"{o}n}}]{si_Pauly2008}%
  \BibitemOpen
  \bibfield  {author} {\bibinfo {author} {\bibfnamefont {F.}~\bibnamefont
  {Pauly}}, \bibinfo {author} {\bibfnamefont {J.~K.}\ \bibnamefont {Viljas}},
  \bibinfo {author} {\bibfnamefont {U.}~\bibnamefont {Huniar}}, \bibinfo
  {author} {\bibfnamefont {M.}~\bibnamefont {H\"{a}fner}}, \bibinfo {author}
  {\bibfnamefont {S.}~\bibnamefont {Wohlthat}}, \bibinfo {author}
  {\bibfnamefont {M.}~\bibnamefont {B\"{u}rkle}}, \bibinfo {author}
  {\bibfnamefont {J.~C.}\ \bibnamefont {Cuevas}}, \ and\ \bibinfo {author}
  {\bibfnamefont {G.}~\bibnamefont {Sch\"{o}n}},\ }\bibfield  {title} {\emph
  {\bibinfo {title} {Cluster-based density-functional approach to quantum
  transport through molecular and atomic contacts},}\ }\href {\doibase
  10.1088/1367-2630/10/12/125019} {\bibfield  {journal} {\bibinfo  {journal}
  {New J. Phys.}\ }\textbf {\bibinfo {volume} {10}},\ \bibinfo {pages} {125019}
  (\bibinfo {year} {2008})}\BibitemShut {NoStop}%
\bibitem [{\citenamefont {B\"{u}rkle}\ \emph {et~al.}(2013)\citenamefont
  {B\"{u}rkle}, \citenamefont {Viljas}, \citenamefont {Hellmuth}, \citenamefont
  {Scheer}, \citenamefont {Weigend}, \citenamefont {Sch\"{o}n},\ and\
  \citenamefont {Pauly}}]{si_Buerkle2013}%
  \BibitemOpen
  \bibfield  {author} {\bibinfo {author} {\bibfnamefont {M.}~\bibnamefont
  {B\"{u}rkle}}, \bibinfo {author} {\bibfnamefont {J.~K.}\ \bibnamefont
  {Viljas}}, \bibinfo {author} {\bibfnamefont {T.~J.}\ \bibnamefont
  {Hellmuth}}, \bibinfo {author} {\bibfnamefont {E.}~\bibnamefont {Scheer}},
  \bibinfo {author} {\bibfnamefont {F.}~\bibnamefont {Weigend}}, \bibinfo
  {author} {\bibfnamefont {G.}~\bibnamefont {Sch\"{o}n}}, \ and\ \bibinfo
  {author} {\bibfnamefont {F.}~\bibnamefont {Pauly}},\ }\bibfield  {title}
  {\emph {\bibinfo {title} {Influence of vibrations on electron transport
  through nanoscale contacts},}\ }\href {\doibase 10.1002/pssb.201350212}
  {\bibfield  {journal} {\bibinfo  {journal} {Phys. Status Solidi B}\ }\textbf
  {\bibinfo {volume} {250}},\ \bibinfo {pages} {2468} (\bibinfo {year}
  {2013})}\BibitemShut {NoStop}%
\end{thebibliography}
\end{document}